\pdfoutput=1

\documentclass[%
 reprint,
 amsmath,amssymb,
 aps,
]{revtex4-2}

\usepackage[T1]{fontenc}
\usepackage[latin9]{inputenc}
\usepackage{textcomp}
\usepackage{mathtools}
\usepackage{amsmath}
\usepackage{amssymb}
\usepackage{graphicx}
\usepackage{url}

\makeatletter

\pdfpageheight\paperheight
\pdfpagewidth\paperwidth

\makeatother

\usepackage{babel}

\begin{document}
\title{Practical hybrid decoding scheme for parity-encoded spin systems}
\author{Yoshihiro Nambu}
\email{y-nambu@aist.go.jp}
\affiliation{NEC-AIST Quantum Technology Cooperative Research Laboratory~~\\
 National Institute of Advanced Industrial Science and Technology }
\begin{abstract}
We propose a practical hybrid decoding scheme for the parity-encoding architecture. This architecture was first introduced by N. Sourlas [Phys. Rev. Lett. \textbf{94}, 070601 (2005)] as a computational technique for tackling hard optimization problems, especially those modeled by spin systems such as the Ising model and spin glasses, and reinvented by W. Lechner, P. Hauke, and P. Zoller [Sci. Adv., \textbf{1}, e1500838 (2015)] to develop quantum annealing devices. We study the specific model, called the SLHZ model, aiming to achieve a near-term quantum annealing device implemented solely through geometrically local spin interactions. Taking account of the close connection between the SLHZ model and a classical low-density-parity-check code, two approaches can be chosen for the decoding: (1) finding the ground state of a spin Hamiltonian derived from the SLHZ model, which can be achieved via stochastic decoders such as a quantum annealer or a classical Monte Carlo sampler; (2) using deterministic decoding techniques for the classical LDPC code, such as belief propagation and bit-flip decoder. The proposed hybrid approach combines the two approaches by applying bit-flip decoding to the readout of the stochastic decoder based on the SLHZ model. We present simulations demonstrating that this approach can reveal the latent potential of the SLHZ model, realizing soft-annealing concept proposed by Sourlas.
\end{abstract}
\keywords{Parity-encoding, error-correcting codes, LDPC, decoding algorithm}
\maketitle

\section{INTRODUCTION}
The spin-glass model is a fundamental concept in physics and mathematics, primarily used to study magnetism and phase transitions in statistical mechanics. A close connection between spin-glass models and combinatorial-optimization problems (COPs) is widely recognized and used as a tool for solving COPs \cite{568530b2-62d5-3d43-9e2f-adfdf424006b}. The key idea is that the energy of the random Ising Hamiltonian can be viewed as a cost function the ground state of which corresponds to a solution of the COP. Probabilistic simulation of Ising spin dynamics, such as simulated annealing (SA) \cite{kirkpatrickOptimizationSimulatedAnnealing1983} and quantum annealing (QA) \cite{kadowakiQuantumAnnealingTransverse1998}, is often used to find the ground state. The spin-glass model has applications in various industrial fields, including routing, scheduling, planning, decision-making, transportation, and telecommunications. A similar connection has been established between the spin-glass model and the classical error-correcting codes (ECCs). Sourlas formulated the decoding of the classical ECCs in terms of Bayesian inference \cite{sourlasSpinglassModelsErrorcorrecting1989}. He showed that the decoding can be expressed as a search for the ground state of the random Ising Hamiltonian. His study implies that decoding classical ECCs can be viewed as a COP, with the associated random Ising Hamiltonian corresponding to the cost function to be minimized (maximized). 

The complexity class of the COPs is known to be NP-hard. When we use the spin-glass models to tackle the COPs, the presence of many local minima separated by large energy barriers is a primary problem. The simulation algorithm gets trapped in those minima. To tackle this problem, Sourlas noted that two mathematically equivalent Hamiltonians can be derived for an ECC: a Hamiltonian given in terms of the original information source and an extended Hamiltonian given in terms of the information source and additional parity information \cite{sourlasSpinGlassesErrorCorrecting1994,sourlasStatisticalMechanicsErrorcorrection1998,sourlasStatisticalMechanicsCapacityapproaching2001}. The extended Hamiltonian involves extra parity-constraint terms. He proposed the soft-annealing concept, which uses the extended Hamiltonian as a new approach to solve hard COPs \cite{sourlasSoftAnnealingNew2005}. The key idea is that one can circumvent barriers and accelerate the dynamics of the algorithm by enlarging the space in which the problem is defined. 

On the other hand, the QA is expected to resolve the above problem through adiabatic quantum computing \cite{RevModPhys.90.015002}. However, to apply a QA device to solve the COP universally, we need to simulate a fully connected graph model in a scalable manner using Ising spin hardware. The soft-annealing approach is also helpful for this purpose. Lechner, Hauke, and Zoller (LHZ) proposed utilizing this approach to embed a random Ising model with all-to-all connectivity within an enlarged Ising system with geometrically local interactions only \cite{lechnerQuantumAnnealingArchitecture2015}. They proposed the parity-encoding (PE) architecture and its concrete realization, which we refer to as the SLHZ model in this paper. In this model, Ising spins are arranged on a plane.  They are connected by four-body interactions in a latticelike manner, which serve as a penalty that constrains the parities of spins. Both the local fields and couplings in the original spin-glass model are mapped to local fields that act on each spin in the SLHZ model. This architecture is particularly attractive for realizing near-term QA devices based on superconducting quantum bits \cite{PhysRevLett.53.1260,PhysRevLett.55.1543,PhysRevLett.55.1908,nakamuraCoherentControlMacroscopic1999} because it is highly compatible with on-chip electronic circuitry \cite{puriQuantumAnnealingAlltoall2017a,niggRobustQuantumOptimizer2017a,zhaoTwoPhotonDrivenKerr2018,onoderaQuantumAnnealerFully2020,yamajiCorrelatedOscillationsKerr2023}.  LHZ also claimed that the SLHZ model is a programmable, scalable, and robust physical platform for QA. Notably, Sourlas was the first researcher to propose soft annealing using the SLHZ model and to recognize that this corresponds to the decoding of associated classical ECCs \cite{sourlasSoftAnnealingNew2005}.

Later, Pastawski and Preskill \cite{pastawskiErrorCorrectionEncoded2016} studied the error-correcting capability of the SLHZ model. They noted that the soft annealing using the SLHZ model can be viewed as decoding of associated classical low-density parity-check (LDPC) code. Based on this observation, they applied the belief-propagation (BP) algorithm, a standard algorithm for LDPC codes \cite{pearlReverendBayesInference1982}, to decode the noisy readout of the SLHZ system \cite{gallagerLowDensityParityCheckCodes1962,gallagerLowDensityParityCheckCodes1963}. They found that one can obtain an error-free state from the noisy readout of the SLHZ model with a high probability if errors are generated by independent and identically distributed (IID) random spin-flip noise.  

To date, however, there has been little investigation focusing on whether the realistic classical decoding technique can fully exploit the potential of the SLHZ model. For example, Albash, Vincl, and Lidar have discussed whether several decoding algorithms may boost the performance of the QA device \cite{albashSimulatedquantumannealingComparisonAlltoall2016}. They have reported that simple majority-vote decoding (MVD) does not improve the performance of QA. Although they have found that more sophisticated minimum-weight decoding (MWD) can boost performance, it is NP-hard and an unrealistic approach for postreadout decoding of QA devices (for details, see Ref. \cite{albashSimulatedquantumannealingComparisonAlltoall2016} and Appendix A). In the context of developing near-term QA, it is preferable to devise a more practical decoding strategy tailored for the SLHZ model that can exploit its potential. 

This paper demonstrates that realistic classical decoding techniques for LDPC codes can unveil potential in the SLHZ model that has not been recognized to date. We propose a practical and straightforward postreadout decoding algorithm tailored to the SLHZ model. The proposed algorithm is an iterative hard-decision decoding algorithm based on majority voting of generalized syndromes, known as Gallager's bit-flipping (BF) algorithm, which has been derived in the context of classical LDPC codes \cite{gallagerLowDensityParityCheckCodes1962,gallagerLowDensityParityCheckCodes1963}. Although the MWD is NP-hard, the BF algorithm is P, similar to the BP algorithms. Moreover, the BF algorithm requires less computational effort and processing time than the BP algorithm and can be implemented in digital logic. To demonstrate the BF algorithm, we have performed two simulations. First, assuming IID spin-flip noise, we have demonstrated that our BF decoding algorithm can recover the noisy output of the SLHZ model, eliminating spin-flip errors with performance comparable to that of the BP algorithm. Second, we have performed classical simulations of stochastically sampled spin readouts from the SLHZ model using a Markov-chain Monte Carlo (MCMC) sampler. This analysis allows us to investigate tolerance to leakage errors arising from dynamical and thermal excitations during the sampling. We show evidence that the BF decoding algorithm can efficiently eliminate spin-flip errors if we make the four-body penalty constraint sufficiently weak. This simulation suggests that a hybrid decoding approach, combining two algorithms---MCMC sampling followed by BF decoding---can be applied to a broader range of spin-flip noise and can mitigate the computational overhead of  SLHZ systems. Our results are consistent with earlier work by Albash et al., who employed simulated quantum annealing to simulate the readout of the QA device based on the SLHZ model (hereafter referred to as the SLHZ-based QA device) when BP decoding was used. We try to understand this result by comparing our algorithm with several extended BF decoding algorithms. According to our analysis, this is a consequence of the combination of the Hamiltonian of the SLHZ system and the properties of BF or BP decoding. We believe that our BF decoding is promising for realizing a near-term SLHZ-based QA device.

This paper is organized as follows. Section \ref{sec:2}  explains the PE architecture, the SLHZ model, and the close connection between classical LDPC codes and the SLHZ model. Section \ref{sec:3} describes classical decoding of the SLHZ model based on Bayesian inference. We propose the BF algorithm as a simple, straightforward decoding algorithm tailored for the SLHZ model. Section \ref{sec:4} demonstrates the proposed BF algorithm under the IID and non-IID spin-flip noise, and shows the performance. In Sec.\ref{sec:5}, we compare several BF decoding algorithms and discuss why our two-stage hybrid decoding strategy tolerates errors originating from non-IID spin-flip noise. We explain why two-stage hybrid decoding outperforms the two algorithms individually. Section \ref{sec:6} concludes this paper. We also discuss the relevance and consistency of our study to the earlier work by Albash et al. \cite{albashSimulatedquantumannealingComparisonAlltoall2016} in Appendix \ref{sec:A} . Supplementary results are also provided in Appendix \ref{sec:B} .

\section{ERROR-CORRECTING CODES AND PARITY-ENCODING ARCHITECTURE\label{sec:2}}

We define the PE architecture and explain the SLHZ model from the viewpoint of classical ECCs, and its connection to classical LDPC codes.

\subsection{Notion}
\begin{figure*}
\includegraphics[viewport=150bp 140bp 800bp 450bp,clip,scale=0.65]{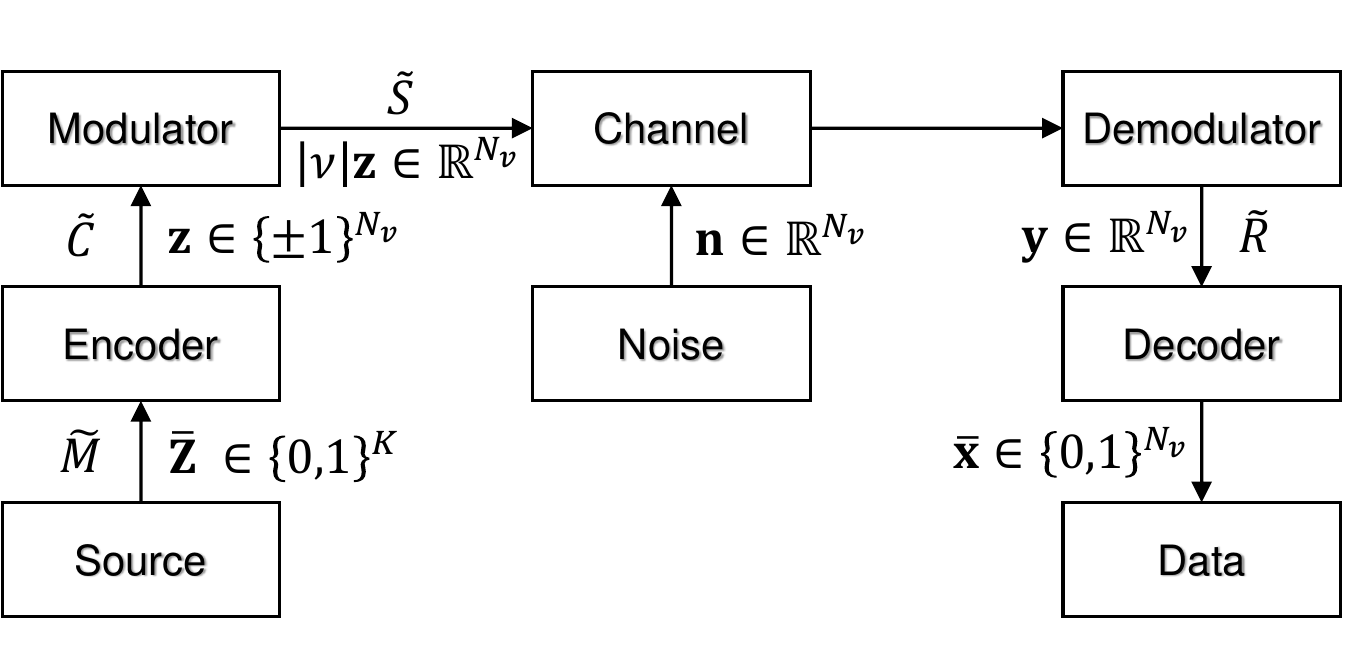}
\caption{A considered model for a communication system.\label{fig:1}}
\end{figure*}
In this paper, we assumed a discrete communication system (Fig.\ref{fig:1}). Suppose that a binary source word $\tilde{M}$ is first encoded into a binary code-word $\tilde{C}$ using some ECC. The code word is modulated into the physical signal $\tilde{S}$ and transmitted. The signal is affected by noise during transmission through a transmission channel. Let $\tilde{M}$ and $\tilde{C}$ be $K$ bits and $N_{v}\left(>K\right)$ bits, respectively, and denote them by the binary vectors $\bar{\textbf{Z}}=(\bar{Z}_{1},\ldots,\bar{Z}_{K})\in\{ 0,1\} ^{K}$ and $\bar{\textbf{z}}=\left(\bar{z}_{1},\ldots,\bar{z}_{N_{v}}\right)\in\{ 0,1\} ^{N_{v}}$, respectively. A linear code is defined by a one-to-one map from the set $\{ \bar{\textbf{Z}}\} $ of $2^{K}$ source words $\tilde{M}$ of length $K$ to the set $\{ \bar{\textbf{z}}\} $ of $2^{K}$ code words $\tilde{C}$ of length $N_{v}$. They are specified by a generating matrix $\textbf{G}_{K\times N_{v}}$ or a generalized parity check matrix $\textbf{H}_{N_{c}\times N_{v}}$ satisfying $\textbf{G}\textbf{H}^{T}=\boldsymbol{0}_{K\times N_{c}}\:\left(\mathrm{mod}\:2\right)$. Here $\textbf{G}$ and $\textbf{H}$ are binary matrices (i.e., their elements are $0$ or $1$) and ``$\mathrm{mod}\:2$'' denotes that the multiplication is modulo two. The source-word $\bar{\textbf{Z}}$ is mapped to the code word $\bar{\textbf{z}}$ by $\bar{\textbf{z}}=\bar{\textbf{Z}}\textbf{G}\:\left(\mathrm{mod}\:2\right)$. Note that any $K$ linearly independent code words can be used to form the generating matrix. For an arbitrary $N_{v}$-dimensional binary vector $\bar{\textbf{x}}=\left(\bar{x}_{1},\ldots,\bar{x}_{N_{v}}\right)\in\left\{ 0,1\right\} ^{N_{v}}$, define the generalized syndrome vector $\bar{\textbf{s}}\left(\bar{\textbf{x}}\right)=\left(\bar{s}_{1}\left(\bar{\textbf{x}}\right),\ldots,\bar{s}_{N_{c}}\left(\bar{\textbf{x}}\right)\right)$ (generalized because it may not have $N_{v}-K$ bits) as $\bar{\textbf{s}}\left(\bar{\textbf{x}}\right)=\bar{\textbf{x}}\textbf{H}^{T}\:\left(\mathrm{mod}\:2\right)$. Then, $\bar{\textbf{x}}$ is a code word if and only if $\bar{\textbf{s}}\left(\bar{\textbf{x}}\right)=\boldsymbol{0}_{1\times N_{c}}$. This vector equation defines a set of generalized parity-check equations, consisting of a set of $N_{c}$ equations where only $N_{v}-K$ ones of them are linearly independent. It follows that we can choose the parity-check matrix for a given linear code in many ways and that we can define many syndrome vectors for the same code. The ratio of the length of the source word to that of the code word is called the rate: $r=K/N_{v}$. Later in this section, we will show specific examples of the matrices $\textbf{G}$ and $\textbf{H}$ for the PE architecture. The code word \textbf{$\bar{\textbf{z}}$} is converted into a sequence of bipolar variables $\bar{\textbf{z}} \rightarrow \textbf{z}=\left(z_{1},\ldots,z_{N_{v}}\right)\in\left\{ \pm1\right\} ^{N_{v}}$ ($0$ is mapped to $+1$, and $1$ to $-1$) and assumed to be modulated to antipodal signals $\left|v\right|\textbf{z}$ by a binary phase shift keying modulation, where $\left|v\right|$ is the signal amplitude. We assume that the signal $\tilde{S}$ is transmitted over a channel subject to additive white Gaussian noise (AWGN). At the end of the channel, the receiver obtains an observation $\tilde{R}$, denoted by an antipodal vector $\textbf{y}=\left(y_{1},\ldots,y_{N_{v}}\right)=\left|v\right|\textbf{z}+\textbf{n}\in\mathbb{R}^{N_{v}}$ , where $\textbf{n}=\left(n_{1},\ldots,n_{N_{v}}\right)\in\mathbb{R}^{N_{v}}$ is a noise vector :: the element of which are IID Gaussian random variables with zero mean and variance $\sigma^{2}$. The purpose of ECC is to communicate reliably over a noisy channel. So, the goal of decoding is to reproduce the original source word $\tilde{M}$ or associated code word $\tilde{C}$ from the observation $\tilde{R}$.

\begin{figure*}[tb]
\includegraphics[viewport=60bp 190bp 870bp 440bp,clip,scale=0.6]{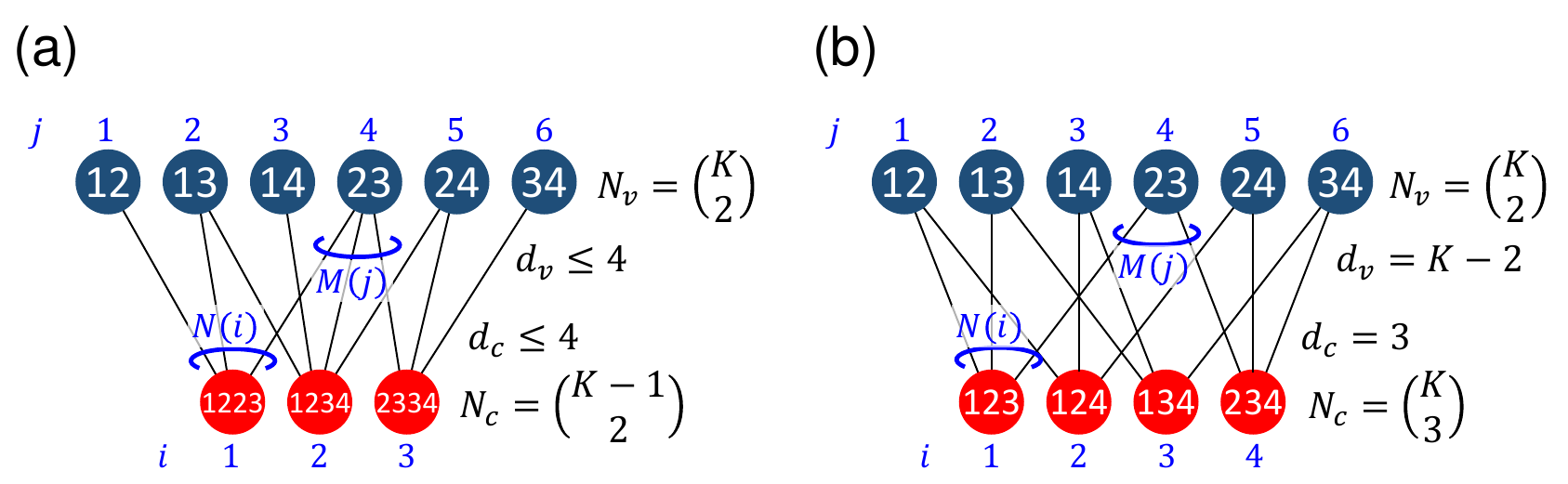}\caption{Two examples of a bipartite graph for $K=4$ logical spins. The dark blue circle with a label $\{k,l\} $ represents the spin variable $x_{kl}$, while the red circles with a label $\{k,l,m,n\}$ or $\{k,l,m\}$ represent the weight-4 syndrome $s_{klmn}^{(4)}$ and weight-3 syndrome $s_{klm}^{(3)}$, respectively. Let us relabel the variables with blue letters. An element of the code-word vector $\textbf{x}=(x_{1},\ldots,x_{N_{v}})\in\{ \pm1\} ^{N_{v}}$ is called a variable node (VN). The $i$ th syndrome of $\textbf{x}$ is defined by $s_{i}(\textbf{x})=\prod_{k\in N(i)}x_{k}\in\{ \pm1\}$ and an element of vector  $\textbf{s}(\textbf{x})=(s_{1}(\textbf{x}),\ldots,s_{N_{c}}(\textbf{x}))\in\{\pm1\} ^{N_{c}}$ is called a check node (CN), wheren the $N(i)=\{ j:H_{ij}(H_{ij}^{'})=1\}$ are the VNs adjacent to a CN $i$ $(1\leq i\leq N_{c})$ and the $M(j)=\{i:H_{ij}(H_{ij}^{'})=1\} $ are the CNs adjacent to a VN $j$ $(1\leq j\leq N_{v})$. The column and row weights of the parity-check matrix are defined by $d_{v}(i)=|M(j)|$ and $d_{c}(i)=|N(i)|$, respectively. Table \ref{table:1} shows the relevant parameters for your reference.
\label{fig:2}}
\end{figure*}

\subsection{Parity-encoding architecture and LDPC codes}

We illustrate a PE architecture with concrete examples. The PE architecture corresponds to the classical ECC according to the following map: $\bar{z}_{ij}=\bar{Z}_{i}\oplus\bar{Z}_{j}$ for $1\leq i\leq j\leq k$. As a simple example, consider the case $K=4$ and assume that $\bar{\textbf{Z}}=\left(\bar{Z}_{1},\bar{Z}_{2},\bar{Z}_{3},\bar{Z}_{4}\right)\in\left\{ 0,1\right\} ^{K}$ and $\bar{\textbf{z}}=\left(\bar{z}_{12},\bar{z}_{13},\bar{z}_{14},\bar{z}_{23},\bar{z}_{24},\bar{z}_{34}\right)\in\left\{ 0,1\right\} ^{\tbinom{K}{2}}$. Because each element of $\bar{\textbf{z}}$ is the binary sum of two elements in $\bar{\textbf{Z}}$, the generating matrix associated with this map is given by the $K\times\tbinom{K}{2}$ matrix: 
\begin{equation}
\textbf{G}=\begin{pmatrix}1 & 1 & 1 & 0 & 0 & 0\\
1 & 0 & 0 & 1 & 1 & 0\\
0 & 1 & 0 & 1 & 0 & 1\\
0 & 0 & 1 & 0 & 1 & 1
\end{pmatrix}.
\end{equation}
Note that there are two instances of 1 in each column in $\textbf{G}$. Then, we can choose the following two parity-check matrices, 
\begin{equation}
\textbf{H}=\begin{pmatrix}1 & 1 & 0 & 1 & 0 & 0\\
0 & 1 & 1 & 1 & 1 & 0\\
0 & 0 & 0 & 1 & 1 & 1
\end{pmatrix}
\end{equation} 
and 
\begin{equation}
\textbf{H}'=\begin{pmatrix}1 & 1 & 0 & 1 & 0 & 0\\
1 & 0 & 1 & 0 & 1 & 0\\
0 & 1 & 1 & 0 & 0 & 1\\
0 & 0 & 0 & 1 & 1 & 1
\end{pmatrix},
\end{equation}
satisfying the constraints $\textbf{G}\textbf{H}^{T}=\textbf{G}\textbf{H}'{}^{T}=\boldsymbol{0}$. From these matrices, two different syndrome vectors can be derived. One is weight-4 syndrome vector $\bar{\textbf{s}}^{(4)}(\bar{\textbf{x}})=\left(\bar{s}_{1223}^{(4)},\bar{s}_{1234}^{(4)},\bar{s}_{2334}^{(4)}\right)=\bar{\textbf{x}}\textbf{H}^{T}\in\left\{ 0,1\right\} ^{\tbinom{K-1}{2}}$, and the other is weight-3 syndrome vector $\bar{\textbf{s}}^{(3)}(\bar{\boldsymbol{x}})=\left(\bar{s}_{123}^{(3)},\bar{s}_{124}^{(3)},\bar{s}_{134}^{(3)},\bar{s}_{234}^{(3)}\right)=\bar{\textbf{x}}\textbf{H}'{}^{T}\in\left\{ 0,1\right\} ^{\tbinom{K}{3}}$, where $\bar{\textbf{x}}=\left(\bar{x}_{12},\bar{x}_{13},\bar{x}_{14},\bar{x}_{23},\bar{x}_{24},\bar{x}_{34}\right)\in\left\{ 0,1\right\} ^{\tbinom{K}{2}}$ is an arbitrary binary vector. $\bar{\textbf{s}}^{(3)}(\bar{\textbf{x}})$ is weight-3 because each of its elements is written as a linear combination of three elements in $\bar{\textbf{x}}$: $\bar{s}_{klm}^{(3)}=\bar{x}_{kl}\oplus\bar{x}_{lm}\oplus\bar{x}_{km}$, where $1\leq k<l<m\leq K$. Similarly, $\bar{\textbf{s}}^{(4)}(\bar{\textbf{x}})$ is weight-4 because each of its elements is written as a linear combination of four elements in $\bar{\textbf{x}}$ if $\bar{\textbf{x}}$ is complemented by the fictitious components $\bar{x}_{ii} \,(i=2,\ldots,K-1)$ assigned by a fixed value $0$: $\bar{s}_{klmn}^{(4)}=\bar{x}_{kl}\oplus\bar{x}_{lm}\oplus\bar{x}_{ln}\oplus\bar{x}_{kn}$.  We note that any element in $\bar{\textbf{s}}^{(4)}(\bar{\textbf{x}})$ can be written as a linear combination of appropriate elements in $\bar{\textbf{s}}^{(3)}(\bar{\textbf{x}})$, and vice versa. 

The connections between the variables $\bar{x}_{kl}$ and the weight-4 syndromes $\bar{s}_{klmn}^{(4)}$ are depicted by a sparse bipartite graph shown in Fig.\ref{fig:2}(a). Similar connections between variables $\bar{x}_{kl}$ and the weight-3 syndromes $\bar{s}_{klm}^{(3)}$ are depicted in Fig.\ref{fig:2}(b). The number of elements in $\bar{\textbf{x}}$ is $N_{v}=\tbinom{K}{2}$, while the numbers of elements in $\bar{\textbf{s}}^{(4)}(\bar{\textbf{x}})$ and $\bar{\textbf{s}}^{(3)}(\bar{\textbf{x}})$ are $N_{c}=\tbinom{K-1}{2}$ and $N_{c}=\tbinom{K}{3}$, respectively. In the terminology of graph theory, the $N_{v}$ elements in $\bar{\textbf{x}}$ constitute variable nodes (VNs) and the $N_{c}$ elements in $\bar{\textbf{s}}^{(4)}(\bar{\textbf{x}})$ or $\bar{\textbf{s}}^{(3)}(\bar{\textbf{x}})$ constitute check nodes (CNs). The matrix $\textbf{H}$ or $\textbf{H}'$ has $N_{c}$ rows and $N_{v}$ columns, where each row $i$ represents a CN and each column $j$ represents a VN; If the entry $H_{ij}=1$, VN $j$ is connected to CN $i$ by an edge. Thus, each edge connecting VN and CN corresponds to an entry $1$ in the row and column of $\textbf{H}$ or $\textbf{H}'$. Let $d_{c}$ be the number of $1$s in each row and $d_{v}$ be the number of $1$s in each column of $\textbf{H}$ or $\textbf{H}'$. These numbers represent the number of edges connected to a VN and a CN, respectively, referred to as row and column weights. The matrix $\textbf{H}'$ is regular, where the row weight $d_{c}=3$ is common for all the CNs and the column weight $d_{v}=K-2$ is common for all the VNs. In contrast, the matrix $\textbf{H}$ is irregular, and its weights depend on the associated nodes, which are at most 4. 

If $K$ is very large, we can always choose a sparse matrix $\textbf{H}$ such that most of the entries are $0$ and a few ones take value $1$. Then, the PE architecture is closely connected to the LDPC codes. Only $2^{K}$ elements out of the $2^{\tbinom{K}{2}}$ possible $\bar{\textbf{x}}$ are valid code words and satisfy the parity-check constraints $\bar{\textbf{s}}^{(4)}(\bar{\textbf{x}})=\boldsymbol{0}_{1\times \tbinom{K-1}{2}}$ or $\bar{\textbf{s}}^{(3)}(\bar{\textbf{x}})=\boldsymbol{0}_{1\times \tbinom{K}{3}}$. Conversely, the syndrome vector  $\bar{\textbf{s}}^{(n)}(\bar{\textbf{x}})$  should be the all-zero vector, i.e., the vector the elements of which are all zero, for the associated vector $\bar{\textbf{x}}$ to be a valid code word. 

\begin{table}[tb]
\begin{tabular}{|c|c|c|c|c|}
\hline 
$K$ & $\tbinom{K}{2}$ & $\tbinom{K-1}{2}$ & $\tbinom{K}{3}$ & $K-2$\tabularnewline
\hline 
\hline 
4 & 6 & 3 & 4 & 2\tabularnewline
\hline 
5 & 10 & 6 & 10 & 3\tabularnewline
\hline 
6 & 15 & 10 & 20 & 4\tabularnewline
\hline 
7 & 21 & 15 & 35 & 5\tabularnewline
\hline 
\end{tabular}\caption{The parameters related to Fig.\ref{fig:2} for $4\leq K\leq7$. \label{table:1}}
\end{table}

\subsection{SLHZ model }

In the remainder of this paper, we will discuss our arguments primarily in the language of spin glasses. Following Sourlas, our argument relies on isomorphism between the additive Boolean group $\left(\left\{ 0,1\right\} ,\oplus\right)$ and the multiplicative Ising group $\left(\left\{ \pm1\right\} ,\cdot\right)$, where a binary variable $\bar{a}{}_{i}\in\left\{ 0,1\right\} $ maps to the spin variable $a_{i}=\left(-1\right)^{\bar{a}_{i}}\in\left\{ \pm1\right\} $ and the binary sum maps to the product by $a_{i}a_{j}=(-1)^{\bar{a}_{i}\oplus\bar{a}_{j}}\in\left\{ \pm1\right\}$ \cite{sourlasSpinGlassesErrorCorrecting1994,sourlasStatisticalMechanicsErrorcorrection1998,sourlasStatisticalMechanicsCapacityapproaching2001}. Source word $\bar{\textbf{Z}}$ and code word $\bar{\textbf{z}}$ are mapped to vectors $\textbf{Z}=\left(Z_{1},\ldots,Z_{K}\right)\in\left\{ \pm1\right\} ^{K}$ and $\textbf{z}=\left(z_{1},\ldots,z_{N_{v}}\right)\in\left\{ \pm1\right\} ^{N_{v}}$ in the spin representation, respectively. Hereafter, we will refer to $\textbf{Z}$ and $\textbf{z}$ as a source-state and a code-state associated with source word $\bar{\textbf{Z}}$ and code word $\bar{\textbf{z}}$, respectively, and associated fictitious spins as logical and physical spins, respectively. Similarly, binary vectors $\bar{\textbf{s}}$ and $\bar{\textbf{x}}$ are mapped to the associated vectors $\textbf{s}$ and $\textbf{x}$ in the spin representation. In the following, we denote variables by symbols with and without an overbar in the binary and spin representations. It should be noted that by isomorphism, every addition of two binary variables corresponds to a unique product of spin variables and vice versa. For example, since $\bar{\textbf{z}}=\bar{\textbf{Z}}\textbf{G}\:\left(\mathrm{mod}\:2\right)$ holds, $\textbf{z}$ and $\textbf{Z}$ are connected by the relation
\begin{equation}
z_{i}=(-1)^{\bigoplus_{j=1}^{K}\bar{Z}_{j}G_{ji}}=\prod_{\{ j:G_{ji}=1\} }Z_{j}\in\{ \pm1\} ,
\label{eq:4}
\end{equation}
where $i=1,\ldots,N_{v}$. Similarly, $\boldsymbol{z}$ must satisfy the equation 
\begin{equation}
(-1)^{\bigoplus_{j=1}^{N_{v}}\bar{z}_{j}H_{ij}}=\prod_{\{ j:H_{ij}=1\} }z_{j}=(-1)^{0}=+1,
\label{eq:5}
\end{equation}
for $i=1,\ldots,N_{c}$, if $\bar{\textbf{z}}$ satisfies the parity-check equation $\bar{\textbf{z}}\textbf{H}^{T}\:\left(\mathrm{mod}\:2\right)=\boldsymbol{0}_{1\times N_{c}}$. 

\begin{figure}[tb]
\includegraphics[viewport=340bp 230bp 580bp 390bp,clip,scale=0.6]{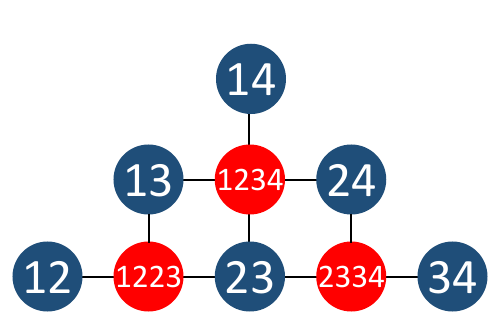}\caption{The bipartite graph that is topologically equivalent to Fig.\ref{fig:2}(a). This graph avoids any edge crossings. 
\label{fig:3}}
\end{figure}

Now, let us consider the graph in Fig.\ref{fig:3}, which is topologically equivalent to Fig.\ref{fig:2}(a). This graph is simply the SLHZ model \cite{lechnerQuantumAnnealingArchitecture2015}. This implies that the SLHZ model is essentially a PE-based model that utilizes the weight-4 syndrome, as viewed from the perspective of the spin-glass model. The plaquette referred to by LHZ \cite{lechnerQuantumAnnealingArchitecture2015} and the stabilizer by Rocchetto, Benjamin, and Li \cite{rocchettoStabilizersDesignTool2016} correspond to the weight-4 syndrome from the perspective of the classical ECCs. Each of the elements is given by a product of four elements in $\textbf{x}$ if $\textbf{x}$ is complemented by the fictitious spin variables  $x_{ii} \,(i=2,\ldots,K-1)$ with fixed value $+1$: $s_{klmn}^{(4)}(\textbf{x})=x_{km}x_{lm}x_{ln}x_{kn}$ \cite{lechnerQuantumAnnealingArchitecture2015}. The vector $\textbf{s}^{(n)}(\textbf{x})$  should be an all-one vector, i.e., the vector the elements of which are all one, for the associated state $\textbf{x}$ of the physical spin to be a valid code state.

\section{POSTREADOUT DECODING OF THE SLHZ MODEL\label{sec:3}}

In this section, we consider postreadout decoding of the SLHZ model from the perspective of the classical ECCs. We first present a general probabilistic framework for classical decoding of ECCs. Next, we propose a simple straightforward decoding algorithm tailored for the SLHZ model.  

\subsection{Probabilistic decoding\label{subsec:2-B}}

Postreadout decoding can be performed according to Bayesian  inference. Consider the conditional probability $P\left(\textbf{z}|\textbf{y}\right)d\textbf{y}$ that the prepared code state is $\textbf{z}$ when the observation was between $\textbf{y}$ and $\textbf{y}+d\textbf{y}$. According to the Bayes theorem,
\begin{equation}
P(\textbf{z}|\textbf{y})=\frac{P(\textbf{y}|\textbf{z})P(\textbf{z})}{\sum_{\textbf{z}}P(\textbf{y}|\textbf{z})P(\textbf{z})}=\kappa P(\textbf{y}|\textbf{z})P(\textbf{z})
\label{eq:6}
\end{equation}
holds, where $\kappa$ is a constant independent of $\textbf{z}$ which is determined by the normalization condition $\mathop{\sum_{\textbf{z}}P\left(\textbf{z}|\textbf{y}\right)=1}$, and $P\left(\textbf{z}\right)$ is the prior probability for the code state $\textbf{z}$. According to Bayesian inference, we can consider the state that maximizes the conditional probability $P\left(\textbf{z}|\textbf{y}\right)$, which offers the most probable word (``word maximum \textit{a posteriori} probability'' or ``word-MAP" decoding). Alternatively, we can consider the state that maximizes its marginals, which offers the most probable symbol (``symbol maximum \textit{a posteriori} probability'' or ``symbol-MAP" decoding). 

\subsubsection*{Word-MAP decoding}

We follow the argument given by Sourlas, who first pointed out the link between the ECCs and the spin glass-model \cite{sourlasSpinglassModelsErrorcorrecting1989}. Sourlas showed that, based on Eqs.(\ref{eq:4}) and (\ref{eq:5}), two different formulations are possible for word-MAP decoding of the same ECC. The first formulation is given in terms of source state $\textbf{Z}$. In this formulation, we assume the following prior probability for $\textbf{z}$: 
\begin{equation}
P(\textbf{z})=\mu\prod_{i=1}^{N_{v}}\delta\left(z_{i},\prod_{\{ j:G_{ji}=1\} }Z_{j}\right),
\label{eq:7}
\end{equation}
where $\mu$ is a normalization constant. The Kronecker's $\delta$'s in Eq.(\ref{eq:7}) enforce the constraint that $\boldsymbol{z}$ obeys Eq.(\ref{eq:4}); it is a valid code state. Assuming that the noise is independent for each spin and that $P\left(\textbf{y}|\textbf{z}\right)=\mathop{\underset{i=1}{\stackrel{N_{v}}{\prod}}P\left(y_{i}|z_{i}\right)}$ holds (memoryless channel), we can derive the following equation:
\begin{eqnarray}
-\ln P(\boldsymbol{z}|\boldsymbol{y})
&=&\mathrm{constant}-\sum_{i=1}^{N_{v}}B_{i}\prod_{\{ j:G_{ji}=1\}}Z_{j}\nonumber\\
&\equiv& H^{source}(\textbf{Z}),
\label{eq:8}
\end{eqnarray}
where 
\begin{equation}
B_{i}=B_{i}(y_{i})=\frac{1}{2}\log\frac{P(y_{i}|z_{i}=+1)}{P(y_{i}|z_{i}=-1)}
\label{eq:9}
\end{equation}
is the half log-likelihood ratio (LLR) for the channel observation $y_{i}$ \cite{sourlasSpinGlassesErrorCorrecting1994}. The vector $\textbf{B}=\left(B_{1},\ldots,B_{N_{v}}\right)\in\mathbb{R}^{N_{v}}$ contains all the information about the observation $\boldsymbol{y}$. In this paper, we focus on the communication through the AWGN channel. The likelihood is given as  
\begin{equation}
P(y_{i}|z_{i})=\frac{1}{\sqrt{2\pi}\sigma}\exp\left[-\frac{(y_{i}-|v|z_{i})^{2}}{2\sigma^{2}}\right],
\label{eq:10}
\end{equation} where $\left|v\right|$ and $\sigma^{2}$ are the amplitude of the prepared signal and the variance of the common Gaussian noise. Then the LLR is given by $B_{i}=\tfrac{1}{2}\beta y_{i}$, where $\beta=\tfrac{2|v|}{\sigma^{2}}>0$ is called the channel reliability factor, and its inverse  $\beta^{-1}$ corresponds to the temperature in the language of spin glasses. Note that $\frac{1}{2}$$\left(\frac{|v|}{\sigma}\right)^{2}$ is the signal-to-noise ratio (SNR) of the AWGN channel. Thus, $B_{i}$ is proportional to the magnitude of the channel observation $y_{i}$ for the AWGN channel. It is evident that $H^{source}(\textbf{Z})$ is in the form of a spin-glass Hamiltonian, where $\tfrac{1}{2}y_{i}$ is identified with the coupling constant $J_{i}$. In this formulation, the word-MAP decoding corresponds to finding the ground state of the Hamiltonian $H^{source}(\textbf{Z})$. 

Alternatively, the second formulation is given in terms of code-state $\textbf{z}$. In this formulation, according to Eq.(\ref{eq:5}), we assume the following prior probability for $\textbf{z}$: 
\begin{equation}
P(\textbf{z})=\mu\prod_{i=1}^{N_{c}} \delta(s_{i}\left(\textbf{z}\right),+1),
\label{eq:11}
\end{equation} where
\begin{equation}
s_{i}(\textbf{z})=\prod_{\{ j:H_{ij}=1\} }z_{j}
\label{eq:12}
\end{equation} is the $i$th syndrome for $\textbf{z}$  in the spin representation. Then, we can derive the following equation: 
\begin{eqnarray}
-\ln P(\textbf{z}|\textbf{y})
&=&-\sum_{i=1}^{N_{v}}B_{i}z_{i}+\underset{\gamma\rightarrow\infty}{\lim}\gamma\sum_{i=1}^{N_{c}}\frac{1-s_{i}(\textbf{z})}{2}\nonumber\\
&\equiv& H^{code}(\textbf{z}).
\label{eq:13}
\end{eqnarray}
In Eq.(\ref{eq:13}), the $\delta$'s in Eq.(\ref{eq:11}) are replaced by a soft constraint using the identity 
\begin{equation}
\delta(x,+1)=\underset{\gamma\rightarrow\infty}{\lim}\exp\left[-\gamma\frac{1-x}{2}\right].
\end{equation}
The second term of Eq.(\ref{eq:13}) enforces the parity constraint that $\textbf{z}$ should obey Eq.(\ref{eq:5}) to be a valid code state. In contrast, the first term captures the correlation between the observation $\textbf{y}$ and the code state $\textbf{z}$. The latter formulation is equivalent to the former one as long as $\textbf{z}$ is a valid code state associated with a source state $\textbf{Z}$, i.e., $s_{i}(\textbf{z})=1$ for any $i$. We note that $H^{code}\left(\textbf{z}\right)$ can be considered a Hamiltonian of an enlarged spin system. In this formulation, the word-MAP decoding corresponds to finding the most probable code state $\textbf{z}$, namely,
\begin{equation}
\textbf{z}=\underset{\textbf{x}\in C}{\arg\max}P(\textbf{x}|\textbf{y})=\underset{\textbf{x}\in C}{\arg\min}H^{code}(\textbf{x}),
\label{eq:15}
\end{equation} 
where $C$ denotes the set of all code states. Such decoded results can be obtained by, e.g., SA and/or QA.

Now, let us recall the SLHZ model. In this case, $s_{i}(\textbf{z})$ is given by the weight-4 syndrome in the spin representation, which is written as a product of four elements in $\boldsymbol{z}$, i.e., $s_{klmn}^{(4)}(\textbf{z})=z_{km}z_{lm}z_{ln}z_{kn}$ \cite{lechnerQuantumAnnealingArchitecture2015}. Then, we see that $H^{code}\left(\textbf{z}\right)$ agrees with the Hamiltonian of the SLHZ model. Therefore, the word-MAP decoding of the LDPC codes is equivalent to finding the ground state of the SLHZ model. Furthermore, since the two formulations, i.e., those based on $H^{code}\left(\textbf{z}\right)$ (Eq.(\ref{eq:13})) and those based on $H^{source}\left(\textbf{Z}\right)$ (Eq.(\ref{eq:8})) are mathematically equivalent, it follows that finding the ground code state of the SLHZ model is equivalent to finding the ground state of the spin glass. 

\subsubsection*{Symbol-MAP decoding}
The above discussion implies that we can solve the COPs through decoding the associated classical LDPC codes. Meanwhile, the word-MAP decoding is not the only strategy for decoding classical LDPC codes. There is another decoding strategy based on different principles. Instead of considering the most probable word, it is also allowed to be interested only in the most probable symbol, i.e., the most probable value $z_{i}$ of the $i$th spin, ignoring the values of the other spin variables. In this strategy, we consider marginals,  
\begin{eqnarray}
P(x_{i}|\textbf{y})
&:=&\sum_{x_{1}=-1}^{+1}\cdots\sum_{x_{i-1}=-1}^{+1}\sum_{x_{i+1}=-1}^{+1}\cdots\sum_{x_{N_{v}}=-1}^{+1}P(\textbf{x}|\textbf{y})\nonumber\\
&=&\sum_{x_{k}(k\neq i)}P(\textbf{x}|\textbf{y}).
\end{eqnarray}
Decoding corresponds to finding the value of the spin variable that maximizes the marginals $P\left(x_{i}|\textbf{y}\right)$, i.e.,
\begin{eqnarray}
z_{i}
&=&\underset{x_{i}\in\{ \pm1\} }{\arg\max}P(x_{i}|\textbf{y})\nonumber\\
&=&\mathrm{sign}[P(x_{i}=+1|\textbf{y})-P(x_{i}=-1|\textbf{y})]\nonumber\\
&=&\mathrm{sign}[L_{i}(\textbf{y}) ],
\label{eq:17}
\end{eqnarray}
where 
\begin{equation}
L_{i}(\textbf{y})=\log\frac{P(x_{i}=+1|\textbf{y})}{P(x_{i}=-1|\textbf{y})}
\end{equation} 
is called the \textit{a posteriori} LLR for the $i$th spin. Since the absolute value $|L_{i}(\textbf{y})|$ gives the error probability of this decision as
\begin{equation}
\epsilon_{i}=\frac{1}{1+e^{|L_{i}(\textbf{y})|}}=\left\{
\begin{matrix}
 P(x_{i}=-1|\textbf{y})&  \lambda_{i}\geq 0\\
 P(x_{i}=+1|\textbf{y})&  \lambda_{i}< 0,\\
\end{matrix}\right.
\label{eq:19}
\end{equation} 
it represents a metric to measure the reliability or uncertainty of a decision in Eq.(\ref{eq:17}), which indicates how likely $x_{i}$ it is to be $+1$ or $-1$. A value close to zero indicates a bad inference, while a larger value indicates a better inference. \cite{masseyThresholdDecoding1962}. 

A variety of algorithms can achieve symbol-MAP decoding. For example, the BP algorithm is widely regarded as the best-known algorithm for LDPC codes \cite{pearlReverendBayesInference1982}. The BP algorithm is an iterative, deterministic algorithm in which the \textit{a posteriori} LLRs, initially given by the observation $\textbf{y}$,  gradually increase in absolute value as messages are iteratively exchanged between the VNs and CNs. 

In the following section, we study an alternative, simple, and straightforward algorithm: the BF algorithm \cite{gallagerLowDensityParityCheckCodes1962,gallagerLowDensityParityCheckCodes1963}. This algorithm is a variety of the symbol-MAP decoding strategies. It uses the syndrome vector $\textbf{s}(\textbf{x})=(s_{1}(\textbf{x}),\ldots,s_{N_{c}}(\textbf{x}))$ as a key to decide the most probable value $z_{i}$ of the $i$th spin, namely, 
\begin{equation}
z_{i}
=\underset{x_{i}\in\{ \pm1\} }{\arg\max}P(x_{i}|\textbf{y})
=\mathrm{sign}[ \tilde{L}_{i}(\textbf{x})],
\end{equation}
where $\textbf{x}=\mathrm{sign}[\textbf{y}]\in\{ \pm1\} ^{N_{v}}$ is the hard decision of the observation $\textbf{y}$ and 
\begin{equation}
\tilde{L}_{i}(\textbf{x})
 =\log\frac
 {P(z_{i}=+1|\boldsymbol{B}_{i}(\textbf{x}))}
 {P(z_{i}=-1|\boldsymbol{B}_{i}(\textbf{x}))}
 \label{eq:21}
\end{equation}
is the associated \textit{a posteriori} LLR for the $i$th spin, where $\textbf{B}_{i}(\textbf{x})$ is the estimator of $z_{i}$ which depends on $\textbf{s}(\textbf{x})$, as shown later. The algorithm starts with an initial decision $\textbf{x}^{(0)}=\mathrm{sign}[\textbf{y}]$, and iteratively updates the decision $x_{i}^{(m)}=\mathrm{sign}[ \tilde{L}_{i}(\textbf{x}^{(m-1)})]$ by a spin-flip operation, which depends on the decision $\textbf{x}^{(m-1)}=(x_{1}^{(m-1)},\ldots,x_{N_{v}}^{(m-1)})$ in the previous round. If the decision $\textbf{x}$ converges after several rounds of iteration, we consider it the most probable code state $\textbf{z}$. 

In the BF algorithm, the reliability of the decision $| \tilde{L}_{i}(\textbf{x})|$ gradually increases as the spin flipping is repeated. It is widely recognized that one of the key advantages of the BF algorithm over the BP algorithm is its simplicity and lower computational complexity. The BF algorithm is particularly useful in environments with limited computational resources because it requires fewer operations to decode the received message. This fact makes it a more efficient option when processing power is a concern. In contrast, although it provides better error-correcting performance in terms of the bit error rate (BER), the BP algorithm typically involves more complex calculations and higher computational requirements. Therefore, the BF decoding is often favored in communication applications where simplicity and low complexity are prioritized.

\subsection{Readout decoding of the SLHZ model }

\subsubsection*{Bit-flipping decoding algorithm}

Now, let us analyze readout decoding of the SLHZ model from the perspective of the LDPC codes. In the following analysis, we assume that the readout $\textbf{r}=(r_{12},\ldots,r_{K-1\,K})\in\left\{ \pm1\right\} ^{\tbinom{K}{2}}$ is available for the $\tbinom{K}{2}$ physical spins in the SLHZ model in accordance with the Ising model. We disregard the information about the observations $\textbf{y}$ of the physical spins, even if it is available in the actual implementation. We will revisit this point later.

We first consider the simplest case, i.e., correcting errors caused by IID noise. This model is appropriate for an SLHZ-based QA device when measurement errors dominate other readout errors. We propose a simple decoding algorithm with very low decoding complexity, which is associated with Gallager's BF decoding algorithm in the spin representation \cite{gallagerLowDensityParityCheckCodes1962,gallagerLowDensityParityCheckCodes1963}. It is based on Massey's \textit{a posteriori} probability (APP) algorithm \cite{masseyThresholdDecoding1962}. Our algorithm only uses information about the syndrome vector $\textbf{s}(\textbf{r})$ to estimate errors $\textbf{e}$ in the current decision $\textbf{r}$ for the readout. Let us introduce the error pattern by $\textbf{e}=\textbf{r}\circ\textbf{z}=(e_{12},\ldots,e_{K-1\,K})\in\left\{ \pm1\right\} ^{\tbinom{K}{2}}$, where $\circ$ denotes component-wise multiplication.  The syndrome vector depends only on $\textbf{e}$ because 
\begin{equation}
\textbf{s}(\textbf{z})=\left(\prod_{\left\{ j:H_{1j}=1\right\} }z_{j},\ldots,\prod_{\left\{ j:H_{N_{c}j}=1\right\} }z_{j}\right)=(+1,\ldots,+1)
\end{equation}
holds for any code state $\textbf{z}$. Therefore, if we note that $z_{ij}\in\left\{ \pm1\right\}$ and $e_{ij}\in\left\{ \pm1\right\}$,  
\begin{equation}
\textbf{s}(\textbf{r})=\textbf{s}(\textbf{z}\circ\textbf{e})=\textbf{s}(\textbf{z})\circ\textbf{s}(\textbf{e})=\textbf{s}(\textbf{e})
\end{equation}
holds. It is important to note that although we never know $\textbf{e}$, we can know $\textbf{s}(\textbf{e})$ from the knowledge of the decision $\textbf{r}$ and utilize it to estimate errors $\textbf{e}$ to be eliminated. 

Estimating the correct value $z_{ij}$ is equivalent to estimating error $e_{ij}$, as $z_{ij}=r_{ij}e_{ij}$ holds. The algorithm uses the weight-3 syndrome $s_{ijk}^{(3)}\left(\textbf{r}\right)=r_{ij}r_{jk}r_{ik}=e_{ij}e_{jk}e_{ik}$ rather than the weight-4 syndrome to decode the current decision $\textbf{r}$ because it has the advantage of treating all variables (VN and CN) symmetrically \cite{pastawskiErrorCorrectionEncoded2016}. The trade-off is that more constraints may increase the decoding complexity. However, as will be discussed later, this is not a fatal problem, as we can leverage symmetry to reduce computational costs. An optimal estimate $e_{ij}^{*}$ can be obtained from $N_{v}=\tbinom{K}{2}$  syndromes $s_{ijk}^{(3)}\left(\textbf{r}\right)$ for $k\neq i,j$. They are considered $K-2$ parity checks orthogonal on $e_{ij}$ \cite{masseyThresholdDecoding1962} and their values suggest whether the parity-check equation is satisfied (being $+1$) or violated (being $-1$). Based on the majority voting, we decide an error variable $e_{ij}^{*}$ whether the spin $ij$  should be flipped ($e_{ij}^{*}=-1$) or not ($e_{ij}^{*}=+1$). The optimal estimate $e_{ij}^{*}$ can be obtained according to the APP decoding by \cite{masseyThresholdDecoding1962} 
\begin{equation}
e_{ij}^{*}
=\mathrm{sign}[\Delta_{ij}(\textbf{r})],
\label{eq:24}
\end{equation}
where 
 \begin{equation}
\Delta_{ij}(\textbf{r})
=w_{0}+\sum_{k\neq i,j}^{K}w_{k}s_{ijk}^{(3)}(\textbf{r})
\label{eq:25}
\end{equation}
is called the inversion function \cite{wadayamaGradientDescentBit2010,sundararajanNoisyGradientDescent2014a}, $w_{0}$ and $w_{k}$ are given by 
\begin{eqnarray}
w_{0}
&=&\log\frac{1-\gamma_{ij}}{\gamma_{ij}},
\\ 
w_{k}
&=&\log\frac{1-p_{k}}{p_{k}},
\end{eqnarray}
with 
\begin{eqnarray}
\gamma_{ij}&=&P(1\neq e_{ij})=P(e_{ij}\neq 1),
\\
p_{k}&=&P( s_{ijk}^{(3)}(\textbf{r})\neq \hat{e}_{ij})=P(e_{jk}e_{ik}\neq 1).
\end{eqnarray}
The decision $r_{ij}^{*}$ can be written as 
\begin{equation}
r_{ij}^{*}
=r_{ij}e_{ij}^{*}
=\mathrm{sign}[L_{ij}^{*}(\textbf{r})],
\end{equation}
where 
\begin{eqnarray}
L_{ij}^{*}(\textbf{r})
&=&r_{ij}\Delta_{ij}(\textbf{r})
=\log\frac
{P(r_{ij}^{*}=+1|\textbf{B}_{ij}(\textbf{r})}
{P(r_{ij}^{*}=-1|\textbf{B}_{ij}(\textbf{r}))},
\\  \nonumber
\textbf{B}_{ij}(\textbf{r})
&=&(B_{ij0}(\textbf{r}),B_{ij1}(\textbf{r}),\ldots,B_{ijK}(\textbf{r}))
\\ 
&=&r_{ij}(1,s_{ij1}^{(3)}(\textbf{r}),\ldots,s_{ijK}^{(3)}(\textbf{r})),
\label{eq:31}
\end{eqnarray}
since 
\begin{eqnarray}
\gamma_{ij}
&=&P(r_{ij}\neq r_{ij}^{*})
=P(B_{ij0}(\textbf{r})\neq r_{ij}^{*}),
\\
p_{k}
&=&P(r_{ij} s_{ijk}^{(3)}(\textbf{r})\neq r_{ij}^{*})
=P(B_{ijk}(\textbf{r})\neq r_{ij}^{*}).
\end{eqnarray}
holds \cite{masseyThresholdDecoding1962,clerkErrorCorrectionCodingDigital1981}. The function $L_{ij}^{*}(\textbf{r})$ can be considered as the LLR for the decision $r_{ij}^{*}$ based on the estimators $\textbf{B}_{ij}(\textbf{r})$, which depends on the weight-3 syndrome vector $\textbf{s}^{(3)}(\textbf{r})$. 
\begin{figure*}
\includegraphics[viewport=150bp 160bp 800bp 400bp,clip,scale=0.65]{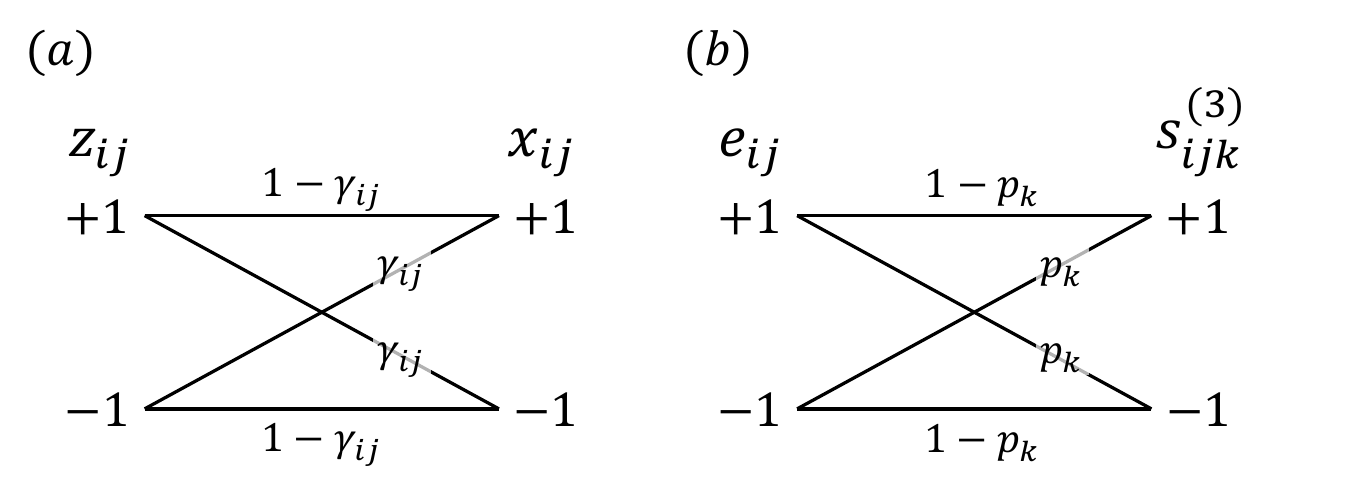}
\caption{The binary symmetric channel associated with the parameters $\gamma_{ij}$ and $p_{k}$.
\label{fig:4}}
\end{figure*}
In Eq.(\ref{eq:24}), $e_{ij}^{*}=-1$ indicates that we should invert the sign of $r_{ij}\rightarrow r_{ij}^{*}=r_{ij}e_{ij}^{*}=-r_{ij}$. Parameters $\gamma_{ij}$ and $p_{k}$ are crosstalk parameters that characterize the binary symmetric channels shown in Fig.\ref{fig:4}. We should note that the probability $p_{k}$ is given by the probability of an odd number of $-1$'s among the errors exclusive of $e_{ij}$ that are checked by $s_{ijk}^{(3)}\left(\textbf{r}\right)=e_{ij}e_{jk}e_{ik}$ so that it is given by \cite{masseyThresholdDecoding1962} 
\begin{equation}
p_{k}=P(e_{jk}e_{ik}=-1)=\frac{1}{2}\left(1-(1-2\gamma_{jk})(1-2\gamma_{ik})\right).
\end{equation}

If we assume IID noise, such that $\gamma_{ij}=\gamma_{0}$ holds for every $\{ i,j\} $ where $0<\gamma_{0}<\tfrac{1}{2}$ , we can confirm that $w_{0}$ can be approximated $w_{0}=w_{k}>0$ for any $k$. Then, we obtain the following simple algorithm: we calculate the best estimate $e_{ij}^{*}$  by 
\begin{equation}
e_{ij}^{*}
=\mathrm{sign}\left(1+\sum_{k\neq i,j}^{K}s_{ijk}^{(3)}(\textbf{r})\right).
\label{eq:36}
\end{equation}
\begin{figure}[tb]
\includegraphics[viewport=290bp 160bp 650bp 390bp,clip,scale=0.7]{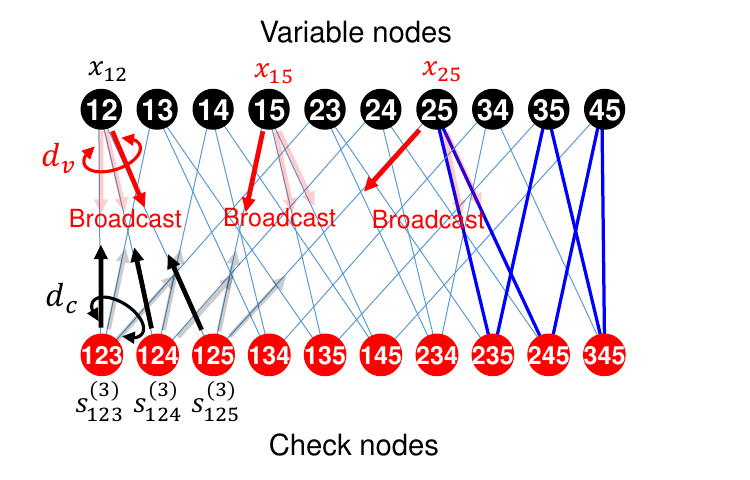}
\caption{The bipartite graph for $K=5$ logical spins. The solid blue lines show an example of the shortest loop of edges connecting VNs and CNs, which has a length of 6.
\label{fig:5}}
\end{figure}
This equation means that if the majority vote of the estimators given by  $\textbf{A}_{ij}(\textbf{r})=\left( 1,s_{ij1}^{(3)}(\textbf{x}),\ldots,s_{ijK}^{(3)}(\textbf{x})\right)$ is negative, $r_{ij}$ should be flipped to increase the value of $\sum_{k\neq i,j}^{K}s_{ijk}^{(3)}(\textbf{r})$. Thus, Eq.(\ref{eq:36}) is reduced to the following best estimate $r_{ij}^{*}\in\left\{ \pm1\right\} $: 
\begin{equation}
r_{ij}^{*}=r_{ij}e_{ij}^{*}=\mathrm{sign}\left(r_{ij}+\sum_{k\neq i,j}^{K}r_{jk}r_{ki}\right).
\label{eq:37}
\end{equation} 
Eq.(\ref{eq:37})  can be interpreted as Gallager's BF decoding \cite{gallagerLowDensityParityCheckCodes1962,gallagerLowDensityParityCheckCodes1963}. Note that the best estimate $r_{ij}^{*}$ is represented in terms of the values of the current variables in VNs. In Fig.\ref{fig:5}, we shows the relevant bipartite graph for $K=5$. Note that the graph associated with the SLHZ system is more loopy, with a minimum length of 4, whereas this graph is less loopy, with a minimum length of 6. The initial decision $\textbf{r}$ on the VNs is broadcast to its adjacent CNs connected by edges. Each CN then reports the syndrome value to its adjacent VNs connected by the edge. All VNs update their values simultaneously according to Eq.(\ref{eq:36}), representing a majority vote of $1$ and the values of adjacent CNs.

\subsubsection*{Application to the SLHZ system}

Next, let us discuss how to apply the BF decoding to the SLHZ model. We introduce a matrix representation of the state of the SLHZ model. Consider the $K\times K$ symmetrized matrix $\hat{\textbf{z}}$, the elements of which are given by spin variable $z_{ij}=z_{ji}\in\left\{ \pm1\right\}$ and have unit diagonal elements, namely, 
\begin{eqnarray}
\hat{\textbf{z}}
&=&\left[\begin{array}{cccccc}
1 & z_{12} & z_{13} & \cdots & z_{1\,K-1} & z_{1\,K}\\
z_{21} & 1 & z_{23} & \cdots & z_{2\,K-1} & z_{2\,K}\\
z_{31} & z_{32} & 1 & \cdots & z_{3\,K-1} & z_{3\,K}\\
\vdots & \vdots & \vdots & \ddots & \cdots & \vdots\\
z_{K-1\,1} & z_{K-1\,2} & z_{K-1\,3} & \cdots & 1 & z_{K-1\,K}\\
z_{K\,1} & z_{K\,2} & z_{K\,3} & \cdots & z_{K\,K-1} & 1
\end{array}\right].
\nonumber
\\
\end{eqnarray}
We use similar notations $\hat{\textbf{r}}$ and $\hat{\textbf{e}}$  for the symmetrized matrices associated with the current readout $\textbf{r}$ and error pattern $\textbf{e}$. They satisfy $\hat{\textbf{r}}=\hat{\textbf{z}}\circ\hat{\textbf{e}}$. Hereafter, we will refer to $\hat{\textbf{e}}$ as the error matrix. Then, Eq.(\ref{eq:37}) can be conveniently written as 
\begin{equation}
\hat{\textbf{r}}^{*}=\mathcal{F}(\hat{\textbf{r}})=\mathrm{sign}\left[\hat{\textbf{r}}(\hat{\textbf{r}}-\textbf{I}_{K\times K})\right],
\label{eq:39}
\end{equation}
where we assume that the sign function is component-wise. This operation updates the $\tbinom{K}{2}$ elements in $\hat{\textbf{r}}$ simultaneously. Therefore, Eq.(\ref{eq:39}) represents a parallel BF algorithm. We further consider the iterative operation of $\mathcal{F}$, i.e., 
\begin{equation}
\hat{\textbf{r}}^{(n)}
=\hat{\textbf{r}}\circ\hat{\textbf{e}}^{(n)}
=\mathcal{F}^{(n)}(\hat{\textbf{r}}).
\label{eq:40}
\end{equation}
If $\hat{\textbf{r}}^{(n)}\rightarrow\hat{\textbf{z}}$, or in other words, $\hat{\textbf{e}}^{(n)}\rightarrow\hat{\textbf{1}}_{K\times K}$ at $n=n_{0}$, where $\hat{\boldsymbol{1}}_{K\times K}$ is $K\times K$ matrix the entries of which are all one (all-one matrix), we say that the decoding has succeeded after $n_{0}$ iterations of the BF decoding. In the case of success, it follows that $s_{ijk}^{(3)}(\hat{\textbf{r}}^{(n)})=1$ for a set of possible $\{ i,j,k\} $ at $n=n_{0}$. Our algorithm can be viewed as an iterative process that gradually increases the number of syndromes with unit values by flipping spins according to the majority vote of the syndromes associated with each spin.

\section{EXPERIMENTAL DEMONSTRATION\label{sec:4}}

We demonstrate that the BF algorithm is valid for eliminating readout errors in the SLHZ model arising not only from IID noise but also from non-IID noise.

\subsection{In the case of IID noise}

\begin{figure*}
\includegraphics[viewport=150bp 20bp 800bp 500bp,clip,scale=0.75]{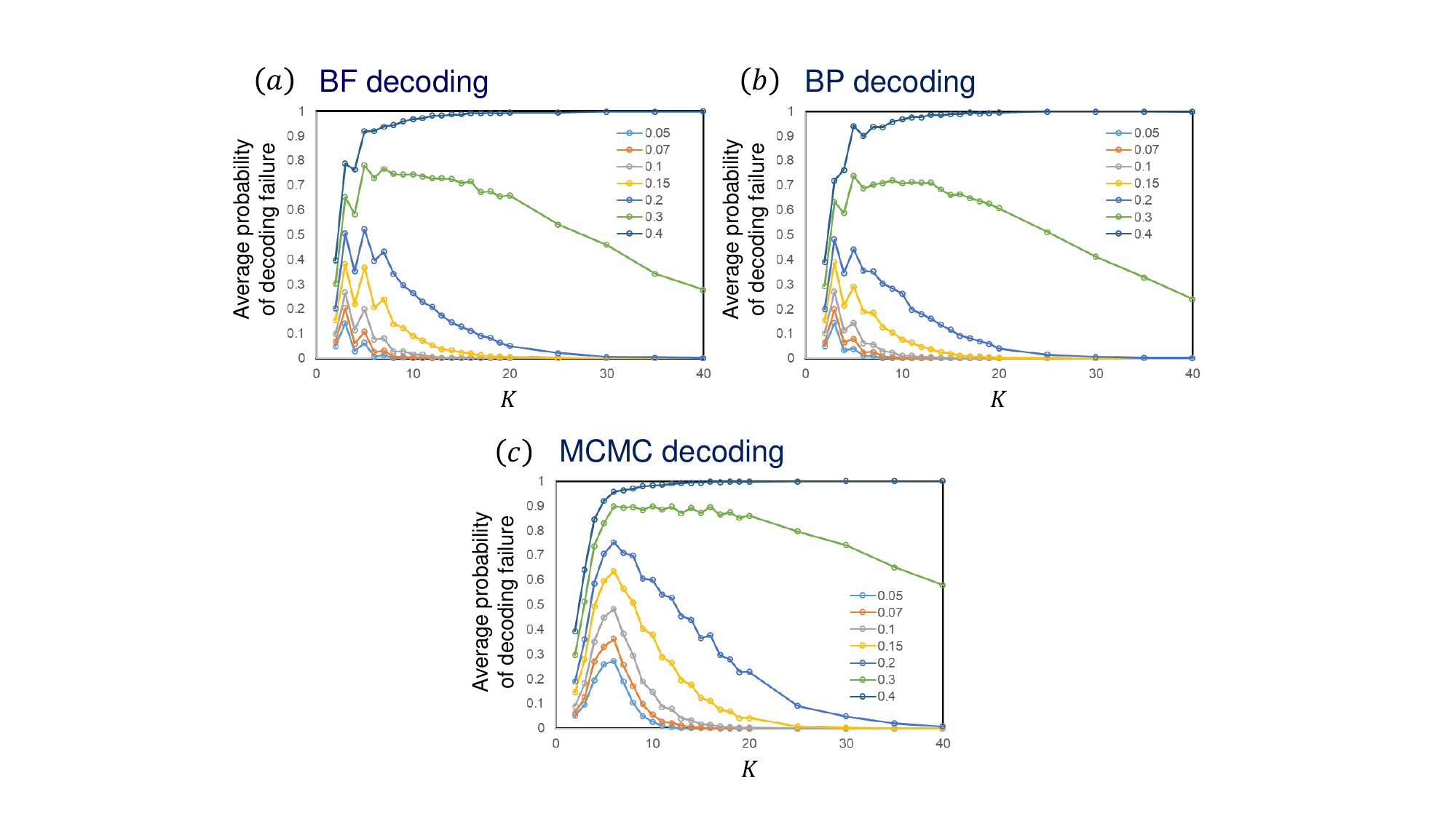}

\caption{A comparison of the performance of the (a) bit-flipping (BF), (b) belief-propagation (BP), and (c) Markov-chain Monte Carlo (MCMC) decodings. Assuming that the error probability $\varepsilon$ is common to all physical spins, the average probabilities of decoding failure are plotted as a function of the number $K$ of logical spins (associated with $\tbinom{K}{2}$ physical spins) for seven values of $\varepsilon$. Each data point has been obtained by averaging over 5000 error-matrix realizations. The BF and BP algorithms have been iterated 5 times for each realization and the MCMC sampling has been iterated $\tbinom{K}{2}$ times. We have considered a tie in the majority voting to be a failure in the BF decoding, although we can resolve it by introducing a tie-breaking rule, such as coin tossing.
\label{fig:6}}
\end{figure*}

\begin{figure*}
\includegraphics[viewport=100bp 220bp 880bp 480bp,clip,scale=0.65]{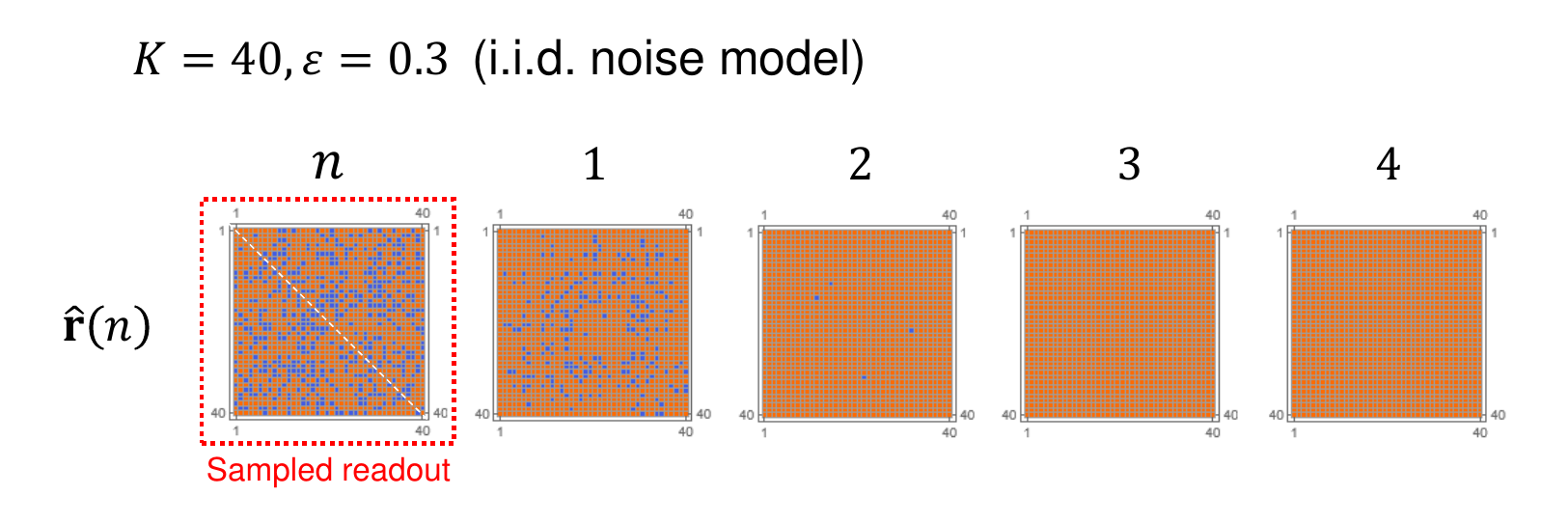}
\caption{A typical example of successfully decoded results by the BF decoding when $K=40$ and $\varepsilon=0.3$. The initial readout has been generated according to the IID-noise model. The estimated matrix $\hat{\textbf{r}}$ is plotted from left to right in increasing order of iterations $n$. The blue pixels represent spins with error.
\label{fig:7}}
\end{figure*}

The performance of the BF algorithm has been investigated under the assumption of IID noise, the simplest model of noisy readout in QA. Pastawski and Preskill have previously studied this model using the BP algorithm \cite{pastawskiErrorCorrectionEncoded2016}. Note that the BF decoding in Eq.(\ref{eq:36}) depends only on the weight-3 syndromes, which treat all variables (VN and CN) symmetrically \cite{pastawskiErrorCorrectionEncoded2016}. Furthermore, if we note that the LDPC codes are linear codes and assume that the AWGN channel is symmetric, i.e. $P(y_{i}|z_{i})=P(-y_{i}|-z_{i})$ holds, the probability of successful decoding is independent of the input code state $\textbf{z}$ \cite{richardsonCapacityLowdensityParitycheck2001b}. In this case, without loss of generality, the performance of decoding can be analyzed using the assumption that the all-one code state $\textbf{z}=\left(+1,\ldots,+1\right)$ has been transmitted \cite{richardsonCapacityLowdensityParitycheck2001b,vuffrayCavityMethodCoding}. In physics, this assumption corresponds to choosing the ferromagnetic gauge and identifying the current decision $\textbf{r}$ with an error pattern $\textbf{e}$. Thus, we assume the input is all-one code-state, that is, $\hat{\textbf{z}}=\hat{\textbf{1}}_{K\times K}$. All the demonstrations have been performed using the \textit{Mathematica}$^{\circledR}$ Ver.14 platform on the Windows 11 operating system. We have generated 5000 symmetric matrices $\hat{\textbf{r}}=\hat{\textbf{e}}$ with unit diagonal elements and other elements randomly assigned $-1$ with probability $\varepsilon<\tfrac{1}{2}$ and $+1$ otherwise. After $n=5$ iterations, we have checked whether the errors have been eliminated. In Fig. \ref{fig:6}(a), we show the performance of the BF algorithm, plotting the probability of decoding failure as a function of the number $K$ of logical spins ranging from $2$ to $40$ (the associated SLHZ model consists of $\tbinom{K}{2}$ physical spins) for seven values of common bit error rate $\varepsilon$ $\left(=0.05,0.07,0.1,0.15,0.2,0.3,0.4\right)$. The MATLAB code implementing the described BF decoder and producing the figure is included as Supplemental Material \cite{*[See Supplemental Material at ] [ for MATLAB code implementation of described BF decoding.] matlabcode}. The failure probability falls steeply as $N$ increases if $\varepsilon$ is not too close to the threshold value $1/2$. A similar performance calculation has been reproduced for the decoding using the BP algorithm after being iterated 5 times, as shown in Fig.\ref{fig:6}(b) \cite{pastawskiErrorCorrectionEncoded2016}. Comparing these figures shows that the performance of the BF algorithm is comparable to that of the BP algorithm. Note that the BP algorithm updates all marginal probabilities $P\left(x_{i}|\textbf{y}\right)$ associated with the $\tbinom{K}{2}$ spins sequentially by passing a real-valued message between the associated VNs and CNs per a single iteration. In contrast, in the BF algorithm, all the $\tbinom{K}{2}$ variables $r_{ij}$ in the matrix $\hat{\textbf{r}}$ are updated in parallel per a single iteration. Thus, each spin variable was updated five times in these algorithms. Fig.\ref{fig:7} is an typical example for a successfully decoded result when $K=40$ and $\varepsilon=0.3$. In this figure, each entry of the error matrix $\hat{\textbf{e}}$ is plotted after $n=1,\ldots,4$ rounds of iteration of the operation in Eq.(\ref{eq:39}). The blue pixels correspond to spins with error ($e_{ij}=-1$), the number of which gradually decreases as rounds of iteration are added. In this example, an error-free matrix has been obtained after $n=3$ iterations. Such results have been observed for more than $70$ \% of the error matrices generated. 

In addition to the BF and BP algorithms, let us focus on the word-MAP decoding introduced in Sec.\ref{subsec:2-B}  and compare its performance with the above two algorithms. The word-MAP decoding says that any code state is the ground state of the Hamiltonian, as shown in Eq.(\ref{eq:13}),  
\begin{equation}
H^{code}(\hat{\textbf{x}})\equiv\gamma\sum_{\{ i,j,k\} }\frac{1-s_{ijk}^{(3)}(\hat{\textbf{x}})}{2},
\label{eq:41}
\end{equation} 
where $\hat{\textbf{x}}\in\{ \pm1\} ^{K\times K}$ is the matrix representing the general state of the SLHZ system. In general, $H^{code}(\hat{\textbf{x}})\geq0$ and $H^{code}(\hat{\textbf{x}})=0$ if and only if $\hat{\textbf{x}}$ is a code-state. In Eq.(\ref{eq:41}), the correlation term (the first term in Eq.(\ref{eq:13})) has been omitted. This is because the all-one code-state assumption uses no information about the observation $\textbf{y}$. Under this assumption, the ground state of $H^{code}(\hat{\textbf{x}})$ is given by 
\begin{equation}
\hat{\textbf{z}}=\underset{\hat{\textbf{x}}\in\{ \pm1\} ^{K\times K}}{\arg\min}H^{code}(\hat{\textbf{x}})=\hat{\boldsymbol{1}}_{K\times K}.
\label{eq:42}
\end{equation}
We have searched the ground state $H^{code}(\hat{\textbf{x}})$ using the classical MCMC sampler, which is equivalent to a word-MAP decoding. We refer to it as the MCMC decoding hereafter. The performance of MCMC decoding has been evaluated based on $H^{code}(\hat{\textbf{x}})$ in Eq.(\ref{eq:41}). We have been generated 5000 random symmetric error matrices $\hat{\textbf{e}}$ with a bit error rate $\varepsilon$. We have used rejection-free MCMC sampling, in which all self-loop transitions are removed from the standard MCMC \cite{nambuRejectionFreeMonteCarlo2022}. We have sampled a sequence $\{ \hat{\textbf{x}}\} $, each of which was of a size $\tbinom{K}{2}$, starting from the initial state $\hat{\textbf{x}}=\hat{\textbf{e}}$. The hyperparameter $\gamma$ has been assigned $\gamma\approx1$, which has been found to be experimentally optimal. We have evaluated the average probability that the ground state $\hat{\textbf{z}}=\hat{\boldsymbol{1}}_{K\times K}$ was found in the sequence of samples $\{ \hat{\textbf{x}}\}$, which gives the success probability of the MCMC decoding. In Fig. \ref{fig:6}(c), we indicate the success probability as a function of $K$ for seven values of $\varepsilon$. Although its performance is not as good as that of the BF and BP algorithms, it shows a similar dependence on $K$ and $\varepsilon$. This result is quite reasonable and suggestive. Later, we will discuss the reason.

\subsection{In the case of non-IID noise}

Next, we consider the problem of primary interest in our study: what is expected when applying the BF algorithm to the readout of the SLHZ-based QA device. As the development of QA devices is ongoing and classical simulation of a quantum mechanical many-body system is computationally challenging, it is difficult to investigate the potential of our BF decoding algorithm when applied to the readouts of an SLHZ-based QA device, both in actual QA devices and in classical computer simulations. In this study, we simulate spin readout in the SLHZ system using data stochastically sampled by a classical MCMC sampler. This analysis enables us to investigate the tolerance of the SLHZ model to leakage errors arising from dynamical and thermal excitations during the sampling.

We have used the following Hamiltonian to sample data from the SLHZ system stochastically: 
\begin{equation}
H^{code}(\hat{\textbf{x}})\equiv-\beta\sum_{\{ i,j\} }J_{ij}x_{ij}+\gamma\sum_{\{ i,j,k,l\} }\frac{1-s_{ijkl}^{(4)}(\hat{\textbf{x}})}{2},
\label{eq:43}
\end{equation}
where $J_{ij}\in\mathbb{R}$ is a coupling constant that is identified with the channel observation $y_{ij}\in\mathbb{R}$ in the AWGN channel model. The first term in Eq. (\ref{eq:43}) corresponds to the distance metric between a code word $\textbf{x}$ and the received information $\textbf{J}$, while the second term corresponds to the constraint satisfaction metric \cite{das2022quantum}. Parameters $\left\{ \beta,\gamma\right\} $ are the Lagrange weights for these two terms. 

If the weight-4 syndrome is defined by 
\begin{equation}
s_{ijkl}^{(4)}(\hat{\textbf{x}})=x_{ik}x_{jk}x_{jl}x_{il},
\end{equation}
with the assumption $x_{ii}=1$ for $i=1,\ldots,K$, Eq.(\ref{eq:43}) is just the Hamiltonian of the SLHZ system given in Fig.\ref{fig:3}. The $(K-1)$-degenerated ground state of the second term in Eq.(\ref{eq:43}) defines the code states. Sampling from the low-temperature equilibrium state of  $H^{code}(\hat{\textbf{x}})$, we can see which is most likely to be the code state given by the matrix $\hat{\textbf{z}}=\textbf{Z}^{T}\textbf{Z}$ that minimizes $H^{code}(\hat{\textbf{x}})$, as shown in Eq.(\ref{eq:42}), where $\textbf{Z}=(Z_{1},\ldots,Z_{K})$ is the source state. Therefore, optimization using this Hamiltonian simply amounts to word-MAP decoding. However, in contrast to Eq.(\ref{eq:42}), $\hat{\textbf{z}}=\hat{\boldsymbol{1}}_{K\times K}$ cannot be assumed in the evaluation of performance in this case. This is because both the correlation term (the first term) as well as the penalty term (the second term) in Eq.(\ref{eq:43}) violate the symmetry conditions required for the all-one code-state assumption to hold. As a result, the performance of the word-MAP decoding based on Eq.(\ref{eq:43}) depends not only on the error distribution $\hat{\boldsymbol{e}}$ but also on $\hat{\boldsymbol{z}}$. This implies that the performance is dependent on a set of coupling constants $\{J_{ij}\}$, and thus, is problem-dependent.

Here, we must draw attention to coupling constants $\{J_{ij}\}$. Recall that the primary purpose of this study is to propose and demonstrate a practical method for readout decoding of the SLHZ-based QA devices. In this context, the $\{J_{ij}\}$ are given by the problem we want to tackle. For example, they are given by an externally defined COP. In other cases, we can consider minimization of the Hamiltonian in Eq,\ref{eq:43} as a decoding of the observation $\textbf{J}=(J_{12},\ldots,J_{K-1\,K})\in\mathbb{R}^{\tbinom{K}{2}}$ in the AWGN channel model. To avoid duplicative consideration in analyzing the potential of the BF algorithm for readout decoding of the SLHZ model, we have disregarded the reliability information for each spin readout of the SLHZ model itself, even if it was available in the actual QA device. Thus, the following results can be considered valid if readout errors are solely due to leakage errors arising from dynamical and thermal excitations during the sampling process, while neglecting other errors, such as measurement errors. In Appendix \ref{sec:B}, we present a supplementary result when IID noise is introduced during the readout stage. This demonstrates that the above simplified approach also exhibits reasonable robustness against IID readout errors.
\begin{figure*}
\includegraphics[viewport=100bp 100bp 840bp 440bp,clip,scale=0.65]{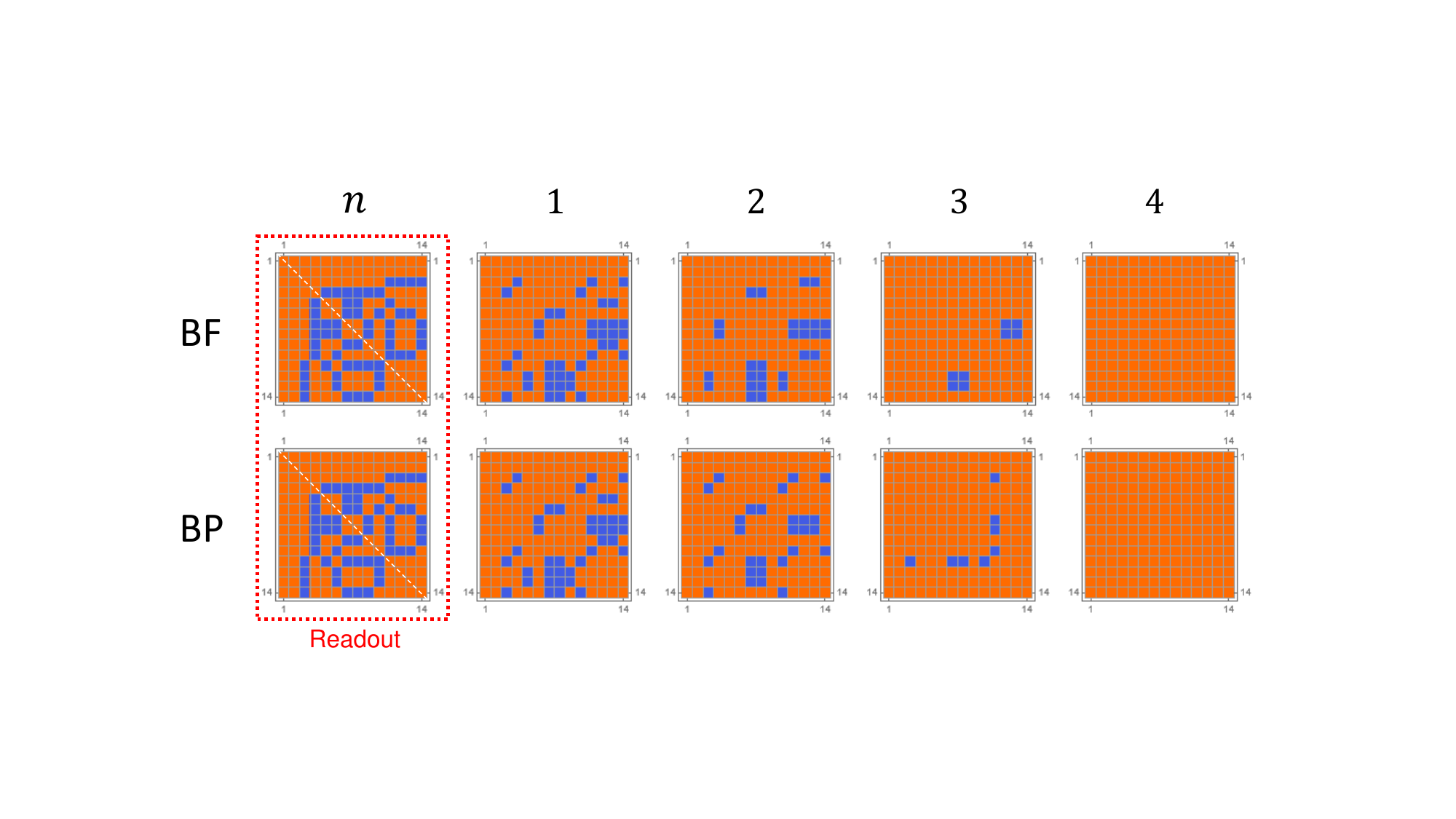}
\caption{Examples of successful BF and BP decoding for $K=14$ logical spins. An MCMC sampler sampled readout based on the Hamiltonian $H^{code}(\hat{\boldsymbol{x}})$ in Eq.(\ref{eq:43}). The readout matrix $\hat{\boldsymbol{r}}$ is plotted from left to right in order of increasing number of iterations $n$ of the two algorithms. The blue pixels represent spins in error.
\label{fig:8}}
\end{figure*}

As a starting point, we provide typical examples demonstrating that both our BF and BP algorithms successfully eliminate errors in the readout of the SLHZ model generated by a classical MCMC sampler. In this example, we have simulated readouts $\hat{\textbf{r}}=\hat{\textbf{e}}\circ\hat{\textbf{z}}\in\{ \pm1\} ^{K\times K}$ of the SLHZ system associated with a spin-glass problem for $K=14$  ($K_{14}$)  as a toy model. We have generated a set of $12$ logical random instances on complete graphs $K_{14}$ with couplings $J_{ij}\in[-\tfrac{1}{4},\tfrac{1}{4}]$ chosen uniformly at random and all logical local fields set to zero, where the code state $\hat{\textbf{z}}$  for a given $\textbf{J}$ has been precomputed by brute force. The leftmost matrix plot in Fig.\ref{fig:8} visualizes a typical example of the error matrix $\hat{\textbf{e}}$ associated with a sampled readout $\hat{\textbf{r}}$ of the SLHZ system. In this example, the error distribution is quite different from the expected one for an IID noise model in Fig.\ref{fig:7}. In Fig.\ref{fig:8}, we show the results where $\hat{\textbf{e}}$ has been decoded by the BF algorithm (upper) and the BP algorithm (lower) when the number of iterations was limited to five rounds. These results suggest that our BF algorithm, as well as the BP algorithm, can eliminate errors in the stochastically sampled readouts of the SLHZ system, which does not follow the IID-noise model. Here, although we cannot know the error rate for each spin because the correct ground state is generally unknown \textit{a priori}, an initial error rate of 0.25 has been assumed for every spin when running the BP algorithm.
\begin{figure*}
\includegraphics[viewport=120bp 40bp 825bp 500bp,clip,scale=0.7]{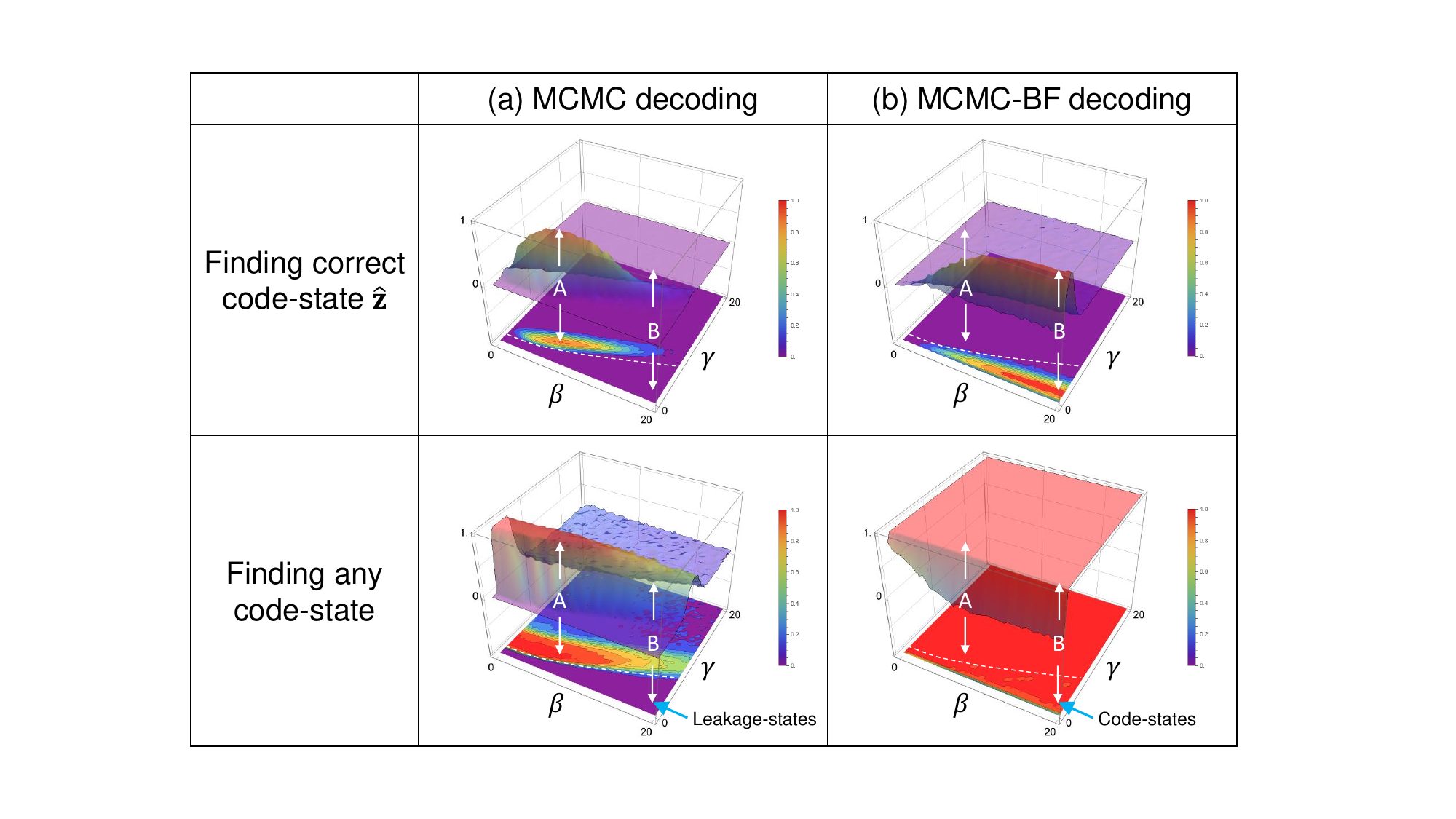}
\caption{Landscapes of the probability distribution for successful decoding. They are plotted as functions of the Lagrange weights $\{ \beta,\gamma\}$ in Eq.(\ref{eq:43}). The left and right columns are the results of (a) MCMC decoding and (b) MCMC-BF hybrid decoding, respectively. The target states in the upper and lower rows of these figures are different: they are the correct code state $\hat{\textbf{z}}$ for the upper row and any code states for the lower row. 
\label{fig:9}}
\end{figure*}
\begin{figure*}
\includegraphics[viewport=120bp 160bp 700bp 450bp,clip,scale=0.75]{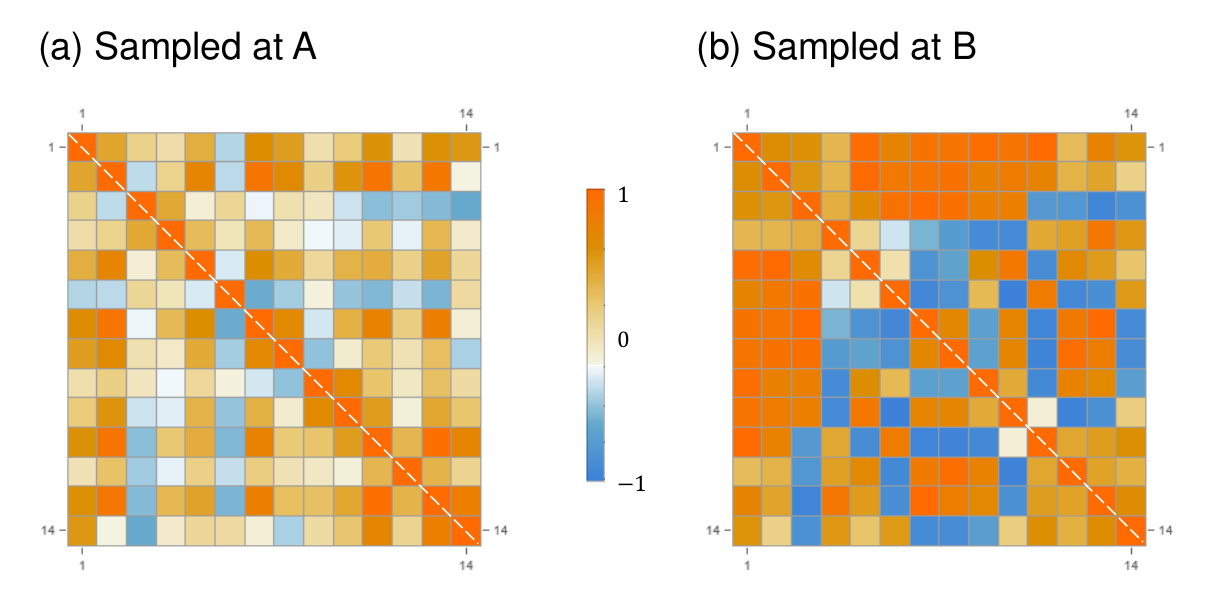}
\caption{Matrix plots representing the averaged error matrix $\langle \hat{\textbf{e}}\rangle =\langle \hat{\textbf{x}}\rangle \circ\hat{\textbf{z}}$ for the sampled states by the MCMC sampling, where each entry reflects the marginal error probability in inferring the correct sign of the associated entry. Warm- (cold-) colored pixels indicate that there is likely no error (error) in the inference of the associated spin. (a), (b) The plots correspond to $\langle \hat{\textbf{e}}\rangle$ when $\{ \hat{\textbf{x}}\} $ has been sampled under the parameter sets (a) $A$ and (b) $B$ in Fig.\ref{fig:9}, respectively. 
\label{fig:10}}
\end{figure*}

The readout in the above examples has been obtained during the evaluation of the following two decoding methods based on $H^{code}(\hat{\textbf{x}})$ in Eq.(\ref{eq:43}). First, we have evaluated the MCMC decoding. Starting from a random state, we have sampled $\hat{\textbf{x}}$ using a rejection-free MCMC sampler and stored a sequence $\{ \hat{\textbf{x}}\}$ by executing MCMC loops many times. The performance has been evaluated by independently repeating the same simulation, obtaining many sampled sequences $\{ \hat{\textbf{x}}\}$. We have evaluated the average probability that $\{ \hat{\textbf{x}}\}$ involves the error-free state $\hat{\textbf{z}}$, which gives the success probability of decoding. In addition to MCMC decoding, we have performed a two-stage decoding, which we refer to as MCMC-BF hybrid decoding. In this method, we have sampled the sequence $\{ \hat{\textbf{x}}\}$ using the same rejection-free MCMC sampler in the first stage. In the second stage, each element of the sequence $\{ \hat{\textbf{x}}\}$ has been decoded using the BF algorithm to eliminate the errors in $\hat{\textbf{x}}$ and update $\{ \hat{\textbf{r}}\} $. Then, we have evaluated the average probability that $\{ \hat{\textbf{r}}\} $ involves $\hat{\textbf{z}}$, which gives the success probability of this hybrid decoding. The MCMC decoding can be identified with simulated annealing. Similarly, the MCMC-BF hybrid decoding can be viewed as a combination of simulated annealing and subsequent classical error correction. 

The performance of the two methods differs significantly. Columns (a) and (b) in Fig.\ref{fig:9} indicate landscapes of the success probability for (a) the MCMC decoding and (b) the MCMC-BF hybrid decoding, respectively. In Fig.\ref{fig:9}, the top row of figures shows the success probability of finding the error-free code state $\hat{\textbf{z}}$, and the bottom row of figures shows the probability for finding any code state $\hat{\textbf{x}}=\textbf{X}^{T}\textbf{X}$, where $\textbf{X}\in\{ \pm1\} ^{K}$ is any logical state. It is very important to note that the sizes of the sample $\{ \hat{\textbf{x}}\}$ for the MCMC decoding and the decoded sample $\{ \hat{\textbf{r}}\}$ for the MCMC-BF hybrid decoding differ significantly; they are $1200\tbinom{K}{2}$ in the MCMC decoding and $4\tbinom{K}{2}$ in the MCMC-BF hybrid decoding. This indicates that the number of iterations required for MCMC sampling is 300 times smaller for the MCMC-BF hybrid decoding than for the MCMC decoding. Therefore, the MCMC-BF hybrid decoding provides a better trade-off between error performance and decoding complexity if the decoding complexity of the BF decoding is negligible. We will revisit the validity of this assumption later. 

Let us note two arrows $A$ and $B$ shown in Fig.\ref{fig:9}. They correspond to the sets of Lagrange weights $A=\{ \beta_{A},\gamma_{A}\}$ and $B=\{ \beta_{B},\gamma_{B}\}$ for which the success probability for decoding is maximized in the MCMC decoding (\textit{A}) and the MCMC-BF hybrid decoding (\textit{B}), respectively. The figure clearly indicates that the parameter sets suitable for use differ between the respective decoding methods. Figures \ref{fig:10}(a) and  \ref{fig:10}(b) are matrix plots showing the average error matrix $\langle \hat{\textbf{e}}\rangle =\langle \hat{\textbf{x}}\rangle \circ\hat{\textbf{z}}$ over the set of samples $\{ \hat{\textbf{x}}\} $ obtained from the MCMC sampler under the parameter sets $A$ and $B$, respectively.  The marginal probability $P_{c}$ that $\hat{x}_{ij}$ gives the correct value, i.e., $\hat{x}_{ij}=\hat{z}_{ij}$, is given by $P_{c}=\tfrac{1+\langle{\hat{e}_{ij}}\rangle }{2}$ . So, a negative entry in $\langle \hat{\textbf{e}}\rangle $ indicates that it is likely that there is an error in the corresponding entry in $\hat{\textbf{x}}$. We can see that $\langle \hat{\textbf{e}}\rangle$ depends on the chosen parameter sets. The average error matrix $\langle \hat{\textbf{e}}\rangle$ in Fig.\ref{fig:10}(b) is consistent with the error matrix $\hat{\textbf{e}}$ in Fig.\ref{fig:8}, because they have been sampled with the same parameter set $B$. Note that it is highly likely to fail sampling a code state by the MCMC sampler for the set $B$ (see the lower left plot in Fig.\ref{fig:9}), while it is highly likely to succeed sampling a code state by the MCMC sampler for the set $A$. This implies that the state $\hat{\textbf{x}}$ sampled in the first stage of the MCMC-BF decoding is not a code state, although the state $\hat{\textbf{x}}$ sampled in the MCMC decoding is a code state. This is quite reasonable because the subsequent BF decoding can correct errors only if $\hat{\textbf{x}}$ sampled at the first stage is a leakage state in the MCMC-BF hybrid decoding. In contrast, the sampled state must be a code state if it is the correct state in the MCMC decoding. This is why the optimal parameter sets $\{ \beta_{opt},\gamma_{opt}\} $ differ between the MCMC decoding and MCMC-BF hybrid decoding. It is important to note that this is a consequence of our BF decoding algorithm and thet it is independent of the sampling algorithm in the first stage. Our results suggest that optimal Lagrange weights should be carefully chosen when applying our BF decoding for readouts of QA, as they may differ from the optimal parameters for QA when used alone. 

\section{DISCUSSION\label{sec:5}}

\subsection{Relationship between MCMC-BF hybrid algorithm and other known BF algorithms\label{subsec:5-A}}

Both the MCMC and BF decoding algorithms are hard-decision algorithms, in which only spin variables are of concern. The MCMC-BF decoding algorithm is considered an extended BF algorithm that incorporates MCMC decoding as a preprocessing step. We discuss the limitations of BF decoding and how they have been resolved in the existing extended BF algorithms. 

There are two obvious limitations in the simple BF decoding algorithm. If we introduce the MCMC decoding as a preprocessing step, these limitations can be overcome. The first limitation is that the BF algorithm neglects the soft information contained in $\textbf{J}$. As mentioned, the BF algorithm assumes a symmetric channel and treats all the spin variables (VNs) and syndromes (CNs) symmetrically, justifying the all-one code-state assumption. This is the result of neglecting the correlation term depending on $\textbf{J}$ , which breaks the necessary symmetry in the Hamiltonian $H^{code}(\hat{\textbf{x}})$ defined by Eq.(\ref{eq:13}) (see also Eq.(\ref{eq:41})). Soft information should be considered to improve decoding, which is frequently fulfilled by taking the reliability information into account in the algorithm. Unfortunately, the BF decoding loses soft information since it approximates the weighted majority vote in Eq. (\ref{eq:24}) and (\ref{eq:25})  to the majority vote in Eq. (\ref{eq:36}) by assuming that $\gamma_{ij}=\gamma_{0}$ for the error probability of every decision $r_{ij}$. In general, $\gamma_{ij}$ may depend on the spin $\left\{ i,j\right\}$. For example, in the AWGN channel model, the LLR for the channel observation $J_{ij}$ is given by $2B_{ij}=\beta J_{ij}$ (see Eq. (\ref{eq:9})). It follows that $\gamma_{ij}$ is small (large) if $\left|J_{ij}\right|$ is small (large) and that the hard decision $x_{ij}=\mathrm{sign}\left[J_{ij}\right]$  is less (more) reliable.

Many researchers have developed sophisticated versions of the weighted BF (WBF) to incorporate reliability information of the hard decisions into the BF algorithm (see details in Refs.\cite{khoaletrungNewDirectionLow,kennedymasundaThresholdBasedMultibit2017}  and references therein). Alternatively, there is another approach to incorporating soft information. For example, Wadayama et al. have proposed the gradient-descent bit-flipping (GDBF) algorithm \cite{wadayamaGradientDescentBit2010}. They are based on modifications to the inversion function that may improve decoding performance, though this comes at the cost of increased decoding complexity. For example, the inversion functions employed in the BF, WBF, and GDBF algorithms are formally written as 
\begin{equation}
\varDelta_{k}^\mathrm{(BF)}(\textbf{x})=1+\sum_{i\in M(k)}s_{i}(\textbf{x}),
\label{eq:45}
\end{equation}
\begin{equation}
\varDelta_{k}^\mathrm{(WBF)}(\textbf{x})=\beta|J_{k}|+\sum_{i\in M(k)}w_{k}s_{i}(\textbf{x}),
\label{eq:46}
\end{equation}
and 
\begin{equation}
\varDelta_{k}^\mathrm{(GDBF)}(\textbf{x})=J_{k}x_{k}+\sum_{i\in M(k)}s_{i}(\textbf{x}),
\label{eq:47}
\end{equation}
respectively. Here, $\textbf{x}=(x_{1},\ldots,x_{N_{v}})\in\{ \pm1\} ^{N_{v}}$ is the current decision for the spin variables (VNs), $\beta$ is a positive real parameter to be adjusted, 
\begin{equation}
s_{i}(\textbf{x})=\prod_{j\in N(i)}x_{j}\in\{ \pm1\}
\end{equation}
is an $i$th syndrome (CN) for the decision $\textbf{x}$ in the spin representation, and $J_{k}$ is identified with channel observation $y_{k}$ in the AWGN channel model. Here, the set $N(i)=\left\{ j:H_{ij}=1\right\} $ are the VNs adjacent to an $i$th CN $\left(1\leq i\leq N_{c}\right)$ and the set $M(j)=\left\{ i:H_{ij}=1\right\} $ is the CNs adjacent to a $j$th VN $\left(1\leq j\leq N_{v}\right)$. For the definition of these set, see Fig.\ref{fig:2}. Note that as long as $s_{i}\left(\textbf{x}\right)$ is a weight-3 syndrome, $\varDelta_{k}^\mathrm{(BF)}(\textbf{x})$ is symmetric for the permutation of the elements of $\textbf{x}$. In other words, $\varDelta_{k}^\mathrm{(BF)}(\textbf{x})$  is invariant under exchanging  $i\longleftrightarrow j$ for any pair of indices of VN other than $k$, justifying the all-one code-state assumption. In contrast, $\varDelta_{k}^\mathrm{(BF)}(\textbf{x})$ is not symmetric if $s_{i}\left(\textbf{x}\right)$ is a weight-4 syndrome. Similarly, because $\varDelta_{k}^\mathrm{(WBF)}(\textbf{x})$  involves weight $w_{k}$ in the second term and $\varDelta_{k}^\mathrm{(GDBF)}(\textbf{x})$ involves $J_{k}$ in the first term, both depend on $\textbf{J}$. Thus,  $\varDelta_{k}^\mathrm{(WBF)}(\textbf{x})$ and $\varDelta_{k}^\mathrm{(GDBF)}(\textbf{x})$ are not symmetric even if $s_{i}(\textbf{x})$ is a weight-3 syndrome. It is interesting to note that the inversion functions in Eq.(\ref{eq:45})-(\ref{eq:47}) can be formally derived from the following Hamiltonians: 
\begin{equation}
H^\mathrm{(BF)}(\textbf{x})=-\sum_{i=1}^{N_{v}}x_{i}-\sum_{i=1}^{N_{c}}s_{i}(\textbf{x}),
\end{equation}
\begin{equation}
H^\mathrm{(WBF)}(\textbf{x})=-\beta\sum_{i=1}^{N_{v}}\left|J_{i}\right|x_{i}-\sum_{i=1}^{N_{c}}w_{i}s_{i}(\textbf{x}),
\end{equation}
\begin{equation}
H^\mathrm{(GDBF)}(\textbf{x})=-\frac{1}{2}\sum_{i=1}^{N_{v}}J_{i}x_{i}-\sum_{i=1}^{N_{c}}s_{i}(\textbf{x}), 
\end{equation}
respectively. If we note that when we invert the sign of $x_{k}$, i.e., $x_{k}\rightarrow-x_{k}$, the increase in energy  $\Delta H_{k}^\mathrm{(X)}(\textbf{x})$ is given by 
\begin{equation}
\Delta H_{k}^{(X)}(\textbf{x})=2\varDelta_{k}^{(X)}(\textbf{x}),
\end{equation}
where $X=\mathrm{BF}$, $\mathrm{WBF}$, or $\mathrm{GDBF}$, it follows that if $\varDelta_{k}^{(X)}(\textbf{x})<0$, flipping the $k$th spin reduces the total energy of the spins. Therefore, the BF, WBF, and GDBF decoding algorithms can be considered deterministic algorithms that determine the most suitable spins to be flipped to reduce the energy $H_{k}^{(X)}(\textbf{x})$ based on the inverse function $\varDelta_{k}^{(X)}(\textbf{x})$.  In addition, in contrast to the BF algorithm, the BP algorithm essentially considers soft information, as it calculates a consistent marginal probability $P\left(x_{i}|\textbf{y}\right)$ by exchanging real-valued messages between VNs and CNs. On the other hand, the inversion function and the associated Hamiltonian for the MCMC decoding are formally written as 
\begin{equation}
\varDelta_{k}^\mathrm{(MCMC)}(\textbf{x})=\beta J_{k}x_{k}+\frac{\gamma}{2}\sum_{i\in M(k)}s_{i}(\textbf{x})
\end{equation}
and
\begin{equation}
H^\mathrm{(MCMC)}(\textbf{x})=-\beta\sum_{i=1}^{N_{v}}J_{i}x_{i}+\gamma\sum_{i=1}^{N_{c}}\frac{1-s_{i}(\textbf{x})}{2}.
\end{equation}
The MCMC decoding algorithm uses MCMC sampling to find $\textbf{x}$ that reduces the energy $H^\mathrm{(MCMC)}(\textbf{x})$ based on the inversion function $\varDelta_{k}^\mathrm{(MCMC)}(\textbf{x})$. Therefore, the MCMC decoding at the first stage inherently incorporates soft information into the correlation term, thereby resolving the limitation of BF decoding.
\begin{figure*}
\includegraphics[viewport=80bp 30bp 840bp 500bp,clip,scale=0.5]{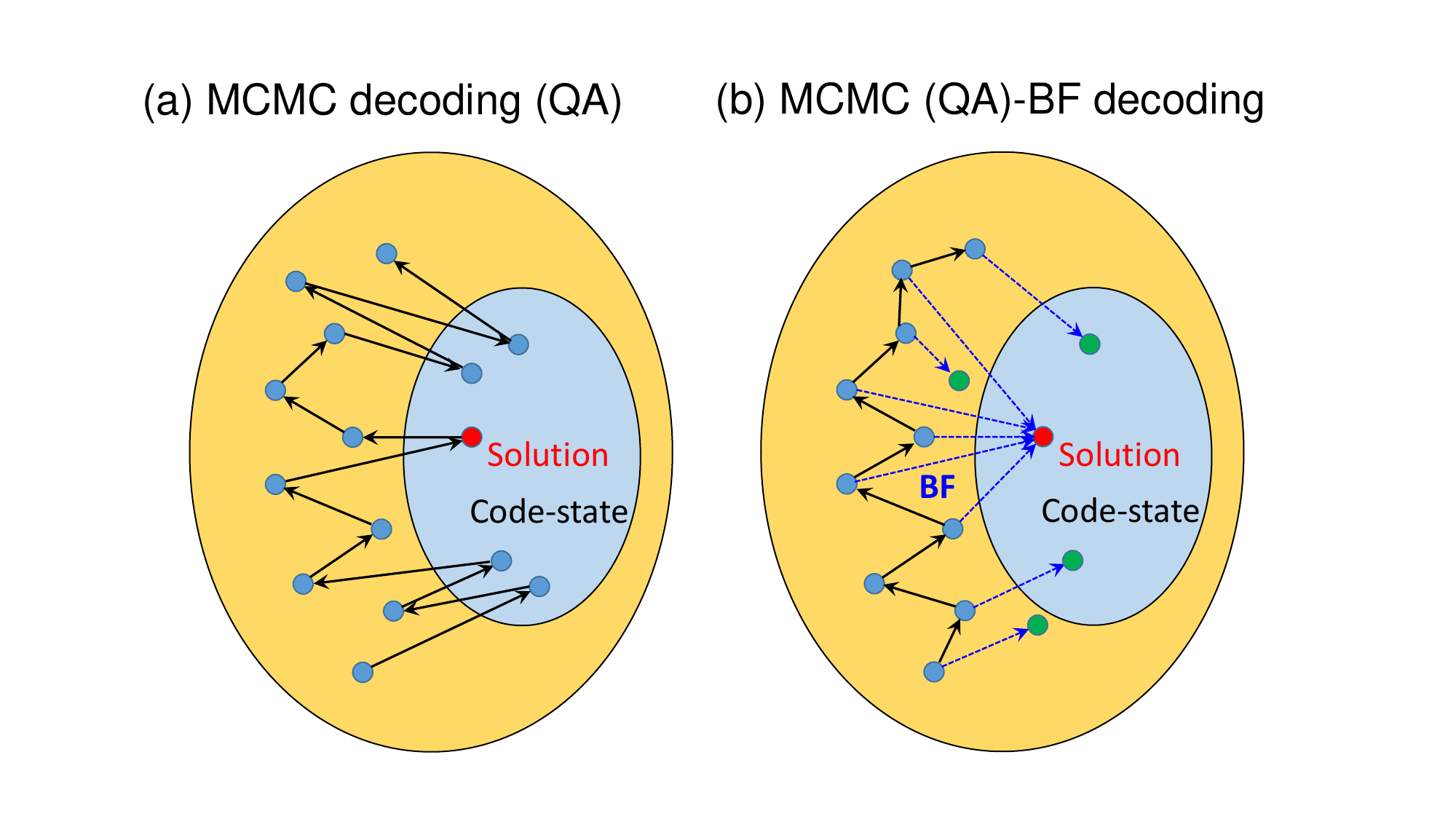}
\caption{Schematic diagrams illustrating two operation modes of the MCMC sampling that optimize (a) MCMC decoding, and (b) MCMC-BF hybrid decoding.  The large and small ellipses indicate the search space and the code-state space, respectively. The blue circles indicate the sampled state by the MCMC sampler. The green circles indicate the decoded states by the BF decoding. The red circle indicates the correct code state $\textbf{z}$. The black solid arrows indicate a sequence of MCMC sampling. The blue broken arrows indicate the map due to the BF decoding.
\label{fig:11}}
\end{figure*}

In addition, MCMC decoding in the first stage also solves another limitation of the BF algorithm in the second stage. The MCMC sampling is stochastic in contrast to our deterministic BF algorithm and the gradient-descent algorithm used in the GDBF algorithm. This stochasticity introduces randomness into the spin-flip selection. It allows spins to flip even when $\varDelta_{k}^\mathrm{(MCMC)}(\textbf{x})>0$, which provides an escape from spurious local minima and makes it more likely to arrive at the neighborhood of the global minimum of $H^\mathrm{(MCMC)}(\textbf{x})$. In this way, the stochastic spin-flip selection helps find the global minimum of $H^\mathrm{(MCMC)}(\textbf{x})$. A similar stochasticity can be incorporated directly into the GDBF algorithm, either by adding a noise term to the inversion function $\varDelta_{k}^\mathrm{(GDBF)}(\textbf{x})$ (noisy GDBF \cite{sundararajanNoisyGradientDescent2014a}) or by taking stochastic spin-flip selection into account in the algorithm (probabilistic GDBF \cite{rasheedFaultTolerantProbabilisticGradientDescent2014}). Briefly speaking, our MCMC-BF hybrid decoding algorithm can be considered an alternative to the BP algorithm and other existing BF decoding algorithms. Our simulation suggests that it is important to properly control the fluctuations of $\textbf{x}$ during MCMC sampling by adjusting the two Lagrange weights $\left\{ \beta,\gamma\right\}$ to optimize the MCMC-BF decoding. It also suggests that the contribution of the penalty term to the Hamiltonian must be small enough not to force $\textbf{x}$ into a valid code state. Controlling Lagrange weights is important for switching between the two decoding modes. In Fig.\ref{fig:11}, we show schematic diagrams illustrating two decoding modes: (a) MCMC decoding and (b) MCMC-BF hybrid decoding, respectively. Decoding is formulated as a constrained COP. The large and small ellipses indicate the search space and the code-state space that minimizes the penalty term in the Hamiltonian. In MCMC decoding, the MCMC samples both the code states (feasible solution) and leakage states (infeasible solution) (Fig.\ref{fig:11}(a)). On some occasions, the correct code state (solution) may be sampled. In MCMC-BF hybrid decoding, on the other hand, the first-stage MCMC sampler samples only the leakage state, and the second-stage BF decoding maps the leakage state to the correct code state occasionally (Fig.\ref{fig:11}(b)). Our simulations have shown that the MCMC-BF hybrid decoding (Fig.\ref{fig:11}(b)) is over 2 orders of magnitude more efficient than the MCMC decoding (Fig.\ref{fig:11}(a)) if we ignore the decoding cost of the postreadout BF decoding and compare performance simply by the number of required MCMC samples needed to obtain at least one correct code state. 

\subsection{Spin-flipping mechanism and decoding cost}

We briefly discuss the mechanism for selecting which spins to flip at each iteration and the decoding cost of the BF algorithm. Spin-flipping mechanisms are free to choose from a range of strategies. For example, we can flip only the most suitable spin chosen based on the inversion function $\varDelta_{k}^{(X)}(\textbf{x})$ values estimated by the current decision $\textbf{x}$. Alternatively, we can flip several spins chosen based on the $\varDelta_{k}^{(X)}(\textbf{x})$ values at once. These are called the single-spin-flipping and multi-spin-flipping strategies, respectively. In this classification, the MCMC decoding algorithm belongs to the single-spin flipping strategy, while our BF decoding algorithm belongs to the multi-spin-flipping strategy. Note that the BF decoding algorithm consists of two operations: evaluating the inversion function $\varDelta_{k}^\mathrm{(BF)}(\textbf{x})$  and determining the spins to be flipped. The inversion function $\varDelta_{k}^\mathrm{(BF)}(\textbf{x})$  for every spin $k$  is computed at once from the current $\textbf{x}$ by matrix multiplication. Subsequent sign evaluation for each element of $\varDelta_{k}^\mathrm{(BF)}(\textbf{x})$ determines which spins are flipped at once.

We can see from Fig.\ref{fig:2} that more edges are connected to VNs for the weight-3 syndromes ($d_{v}=K-2$) than for the weight-4 syndromes ($d_{v}\leq4$). Thus, each spin variable affects more syndromes when using the weight-3 syndrome than the weight-4 syndrome. Conversely, each spin variable is determined from more syndromes when using the weight-3 syndrome than the weight-4 syndrome. Consequently, when using the weight-3 syndrome, one must compute the sum of products of data distributed globally across many spins to decide whether to invert each spin. In contrast, when using the weight-4 syndrome, it is sufficient to compute the sum of products of data distributed over at most eight adjacent spins to determine whether to invert each spin. The reader may fear that a global reduction operation to compute the inversion function when using weight-3 syndrome will require more effort at the expense of  accelerating error correction at each spin flip, but this is not the case, as discussed below. 

We can perform the matrix multiplication required for calculating $\varDelta_{k}^\mathrm{(BF)}(\textbf{x})$ easily and gain significant benefits from parallel-computing techniques and hardware engines that compute matrix sum products, such as GPUs and vector processing engines developed for machine learning. This fact provides the practical merit of choosing the weight-3 syndrome in $\varDelta_{k}^\mathrm{(BF)}(\textbf{x})$ (Eq.(\ref{eq:45})). In fact, comparing the computation times of the MCMC and MCMC-BF hybrid decoding presented in Fig.\ref{fig:9} performed on the \textit{Mathematica}$^{\circledR}$ platform, the computation time of MCMC-BF hybrid decoding was one fifth of that of the MCMC decoding. Although the computational cost depends on the algorithm used and the software or hardware platforms on which we perform the calculations, this result suggests that the MCMC-BF hybrid decoding is more efficient than MCMC decoding alone. Furthermore, we confirmed that our BF algorithm was much less computationally expensive than the BP algorithm. For example,  the computation time of the performance evaluation for the BF decoding shown in Fig.\ref{fig:6} (a) was about 1/40 of that for the BP decoding shown in Fig.\ref{fig:6}  (b). 

This paper does not discuss decoding cost in any further detail. Let us recall that the purpose of this paper is to propose the BF decoding and demonstrate its potential as a practical postprocessing algorithm for a preprocessing stochastic algorithm, such as a QA. Since the development of QA devices is ongoing, it is not easy to demonstrate our BF decoding algorithm as a postprocessing step for an actual QA device. Instead, the potential of the BF decoding has been demonstrated using a classical MCMC sampler as a preprocessor in this study. We believe that the present result depends primarily on the high compatibility between the SLHZ model and the BF decoding algorithm, rather than on the MCMC sampler. Furthermore, our result is consistent with the general belief that two decoding algorithms based on different mechanisms can be combined to solve problems that neither can solve alone. We believe that our BF decoding algorithm is promising for correcting readout errors of the QA devices due to measurement errors as well as dynamic errors that may be encountered during QA. 

\section{CONCLUSIONS\label{sec:6}}

This paper has proposed a practical decoding algorithm to correct readout errors tailored for the SLHZ model. Given the close connection between the SLHZ model and classical LDPC codes, classical decoding techniques for LDPC codes are expected to help exploit the latent potential of the SLHZ model. We have proposed a simple BF algorithm based on iterative majority voting, a decoding algorithm for LDPC codes, which is much simpler than the standard BP algorithm. We have demonstrated that the BF algorithm provides strong protection against IID noise for spin readouts of the SLHZ system and is as efficient as the BP algorithm. To study the tolerance to errors caused by a broader range of noise, we have conducted a classical simulation of spin readouts on the SLHZ system using the MCMC sampler. We have found that BF decoding can correct leakage errors caused by dynamical and thermal excitations during MCMC sampling. Efficient decoding is possible if we reduce energy penalties, allowing thermal excitations to populate correctable leakage states more frequently. This is quite reasonable, given that stochastic sampling followed by classical error correction can be regarded as a two-stage hybrid decoding algorithm. Our observation suggests that the SLHZ model exhibits error correction against a broader range of noise models than the IID-noise model when combined with postreadout BF decoding. Note that the BF algorithm does not suffer from the energy-barrier issue inherent to stochastic sampling. Thus, the two-stage hybrid decoding algorithm can be regarded as a practical implementation of the soft-annealing proposed by Sourlas \cite{sourlasSoftAnnealingNew2005}. Controlling the Lagrange weights in the first-stage sampler is crucial for efficiently obtaining the correct state. Special attention should be paid to the fact that the optimal Lagrange weights depend heavily on whether postreadout decoding is performed.

Since our demonstration has used readouts sampled stochastically by a classical method, specifically MCMC sampling, we are not necessarily convinced that the BF decoding is valid for readouts obtained through an SLHZ-based QA device. Further research is needed to assess how effectively the QA device can collaborate with BF decoding under realistic conditions. Nevertheless, we believe that most of our insights stem from the intrinsic nature of the SLHZ model and BF decoding and are applicable regardless of the stochastic sampling mechanism employed in the first stage. For example, if measurement error is the primary source of readout error of the SLHZ system, our BF decoding algorithm offers a straightforward solution to mitigate it. In addition, it would be reasonable to expect that a two-stage hybrid computation combining QA and postreadout BF decoding may address issues that neither method can solve independently. Our research also emphasizes the importance of selecting appropriate decoding algorithms to exploit the potential of SLHZ systems. 

In this study, we have tested only a small number of instances $K_{14}$ to investigate the performance and characteristics of the BF decoding; however, this is insufficient to draw a general conclusion. Of particular importance for future study is to investigate the size dependence of the SLHZ model on the performance. In our discussion, we have suggested that post-readout BF decoding may partially compensate for the poor spin-update properties of the SLHZ system (for details, see the Appendix \ref{sec:A}). In particular, it will be interesting to see whether this compensation works well for a problem of arbitrary size. Albash et al. have also pointed out another significant limitation of the SLHZ system: the strength of the energy penalty must grow with the problem size \cite{albashSimulatedquantumannealingComparisonAlltoall2016}. Introducing postreadout BF decoding may alleviate this limitation by reducing the penalty strength. Both these issues are important for achieving scalability. More research is needed to determine the efficacy of postreadout decoding. 

\begin{acknowledgments}
I would like to thank Dr. T. Kadowaki at the Global Research and Development Center for Business by Quantum-AI technology (G-QuAT) and Professor H. Nishimori at the Institute of Science Tokyo for their valuable comments and discussions. I also thank Dr. Masayuki Shirane of NEC Corporation and the National Institute of Advanced Industrial Science and Technology for his continuous support. This paper is partly based on results obtained from a project, Grant No. JPNP16007, commissioned by the New Energy and Industrial Technology Development Organization (NEDO), Japan. 
\end{acknowledgments}

\appendix
\section{RELEVANCE AND CONSISTENCY TO THE EARLIER WORK BY ALBASH et al.}\label{sec:A}

In Ref.\cite{albashSimulatedquantumannealingComparisonAlltoall2016}, the authors have present minimum-weight decoding (MWD), which aims to find an error pattern $\textbf{e}$ with minimum Hamming distance from the hard-decided readouts $\textbf{r}$ and with the same syndrome vector as $\textbf{s}^{(4)}(\textbf{r})$. They have directly estimated  the optimal error pattern $\textbf{e}^{*}\in\{\pm1\}^{N_v}$ from the global ground state of the following Hamiltonian: 
\begin{equation}
H^\mathrm{(MWD)}(\textbf{x})
=-\sum_{i=1}^{N_{v}}x_{i}-\lambda \sum_{i=1}^{N_{c}} s_{i}^{(4)}(\textbf{r}) s_{i}^{(4)}(\textbf{x}),
\label{eq:55}
\end{equation}
where $\textbf{x}=(x_{1},\ldots,x_{N_{v}})$  is an arbitrary spin state and $\textbf{s}^{(4)}(\textbf{x})=(s_{1}^{(4)}(\textbf{x}),\ldots,s_{N_{c}}^{(4)}(\textbf{x}))$  is the associated weight-4 syndrome vector. Here, $\lambda$ is assumed to be sufficiently large such that $\textbf{x}$ minimizes all constraint terms. The second term in Eq.(\ref{eq:55})  forces $\textbf{x}$ to satisfy $\textbf{s}^{(4)}(\textbf{x})=\textbf{s}^{(4)}(\textbf{r})$ and the first term minimizes the number of spins with $x_{i}=-1$. Let $C$ denote the set of $\textbf{x}$ satisfying $\textbf{s}^{(4)}(\textbf{x})=\textbf{s}^{(4)}(\textbf{r})$. Then, the optimal estimate $\textbf{e}^{*}$ is given by
\begin{equation}
\textbf{e}^{*}=\underset{\textbf{x}\in C}{\arg\min}H^\mathrm{(MWD)}(\textbf{x}).
\end{equation}
Thus, $\textbf{e}^{*}$ is the error pattern with the same syndrome vector and the smallest number of errors (i.e., the smallest number of elements $e_{k}$ with a value $-1$). Consequently,  the state $\textbf{z}^{*}$ estimated by the MWD is given by $\textbf{z}^{*}=\textbf{r}\circ\textbf{e}^{*}$. The authors have also related the MWD to finding the ground state of the following spin-glass Hamiltonian: 
\begin{equation}
H_\mathrm{spin}^\mathrm{(MWD)}(\textbf{s})=-\sum_{i<j}g_{ij}s_{i}s_{j},
\label{eq:57}
\end{equation}
where a set of spin variables $\textbf{g}=(g_{12},\ldots,g_{K-1\,K})\in\{\pm 1\}^{\tbinom{K}{2}}$ can be regarded as a gauge transformation. Once the ground state of Eq.(\ref{eq:57}) is found, one can deduce the optimal estimate  $\textbf{e}^{*}$ and obtain the estimate $\textbf{z}^{*}$. The problem here is that the MWD requires the solution that minimizes the Hamiltonian defined in Eq. (\ref{eq:57}). This is equivalent to solving an instance of maximum 2-satisfiability, which is as hard as the original COP to be solved. Therefore, the MWD does not necessarily resolve the performance bottleneck for the SLHZ system. We think that MWD is challenging to apply in realistic scenarios when considering implementation in the near-term QA devices. In contrast, our BF decoding algorithm is an approximation, in which the current estimate $\textbf{r}$ is iteratively updated to approach a better estimate $\textbf{r}'$. If the estimate $\textbf{r}$ converges to a certain state within finite iterations, it gives a correct estimate $\textbf{z}^{*}$ with a finite probability. Our BF algorithm can be executed on a deterministic Turing machine in polynomial time, since it completes with a finite number of iterations of matrix multiplication and the associated sign evaluation of every entry of the resultant matrix. Therefore, we believe that our BF decoding algorithm is practical and realistic for implementation in near-term QA devices.

The authors have also pointed out that the SLHZ system faces more challenges with single spin updates than the other embedding method, i.e., the minor-embedding (ME) model \cite{choiMinorembeddingAdiabaticQuantum2008a,choiMinorembeddingAdiabaticQuantum2011a}. We expect that such challenges can be alleviated by introducing the BF decoding. Recall that, in the ME and SLHZ models, there is an overhead resulting from the embedding, where the chain of short-range physical interactions simulates the effect of long-range physical interactions. A chain consists of local two-body interactions in the ME model, while it consists of local four-body interactions in the SLHZ model. On the other hand, the spin update in our BF algorithm relies on numerous weight-3 syndromes, including long-range three-body interactions. As shown in Fig.\ref{fig:2}, the column weight $d_{v}$ for the weight-4 syndrome is always less than 4, while that for the weight-3 syndrome depends on the size $K$ of the original logical problem and increases with increasing $K$. For $K>6$, the use of the weight-3 syndrome instead of the weight-4 syndrome can make spin updates more efficient since it increases the connectivity between spins. This has actually been confirmed in the following observation.  Let us recall that MCMC decoding using the Hamiltonian in Eq.\ref{eq:41} results in the performance curves shown in Fig \ref{fig:6}(c). However, when a similar evaluation has been performed by replacing the weight-3 syndrome $s_{ijk}^{(3)}(\hat{\textbf{x}})$ with the weight-4 syndrome $s_{ijk}^{(4)}(\hat{\textbf{x}})$, the performance has been found to be much worse. This is quite reasonable, considering that the mixing property of the MCMC depends on connectivity between the spins; the more connectivity each spin has, the better the mixing property should be. In other words, embedding is not possible without sacrificing mixing properties. Since we calculate BF decoding on a digital computer, we do not have to worry about connectivity issues. Therefore, there is no reason to avoid using weight-3 syndromes in the postreadout BF decoding. Even if the spin-update of the SLHZ system is less efficient, the postreadout BF decoding can vastly recover its unfavorable spin-update properties by using weight-3 syndromes. 

Our simple BF algorithm fails to account for the soft information incorporated in the logical coupling constants $\textbf{J}=(J_{12},\ldots,J_{K-1\,K})\in\mathbb{R}^{\tbinom{K}{2}}$. Thus, it is required to increase the relative importance of the correlation terms to the four-body penalty terms to incorporate the soft information contained in the Hamiltonian $H^{code}\left(\hat{\textbf{x}}\right)$ of the SLHZ model. Previously, Albash et al. had tested using distributions sampled by simulated quantum annealing (SQA) and parallel tempering (PT). They found that BP, as well as MWD, offers a substantial performance boost over MVD when the penalty strength is brought close to zero. However, they concluded that this boost originated from MWD itself, because it was typically observed in cases where the ground state of the original logical problem and the same problem approximated by $g_{ij}=\mathrm{sign}(J_{ij})$ happen to coincide. In our simulation, a similar performance boost has been observed for the BF decoding when the penalty strength was almost zero, as shown in Fig.\ref{fig:9}. Moreover, a performance boost has been observed in all instances studied, regardless of whether they belonged to the above specific case. We believe that this performance boost comes from the mechanism schematically depicted by Fig.\ref{fig:11}  in Sec. \ref{subsec:5-A}. The readout state just before decoding by our BF algorithm must be a state with correctable leakage errors. By reducing the penalty strength, the soft information contained in the correlation term of the Hamiltonian $H^{code}\left(\hat{\textbf{x}}\right)$ of the SLHZ system can be better reflected, thereby boosting the sampling probability of states with correctable leakage errors. As shown above, our study suggests that the BF and BP algorithms can cope not only with IID errors but also with a broader range of errors. 

We recognize that the performance boost of the MCMC-BF hybrid decoding at a vanishing penalty strength provides evidence supporting the inherent error-correcting capability of the SLHZ model. Our study has suggested that controlling the penalty strength is crucial to exploiting the potential of the SLHZ model. Nonetheless, we do not claim that this performance boost is sufficient for the SLHZ model to outperform the ME model. Rather, we consider it essential that the SLHZ system be competitive with the ME model. For supplementary information, we have confirmed in our preliminary experiments that the ME model shows no performance boost after postreadout MVD. 

\section{SUPPLEMENTARY RESULT ON EXPERIMENTAL DEMONSTRATION}\label{sec:B}
In section \ref{sec:4}, we have demonstrated that the postreadout BF decoding algorithm can eliminate not only IID errors but also leakage errors due to dynamic and thermal excitations during the MCMC sampling process. However, when we imagine that the BF decoding is applied to the readout of actual QA devices rather than MCMC simulation, several types of errors may occur simultaneously. In this appendix, we assume IID errors in the readouts of the MCMC sampler which might be a realistic model of readout errors for imperfect annealing devices, and investigate their impact on the performance of the MCMC-BF hybrid decoding. We present a result showing that the MCMC-BF decoder is robust to such readout errors to some extent.
\begin{figure}[tb]
\includegraphics[viewport=320bp 160bp 650bp 380bp,clip,scale=0.75]{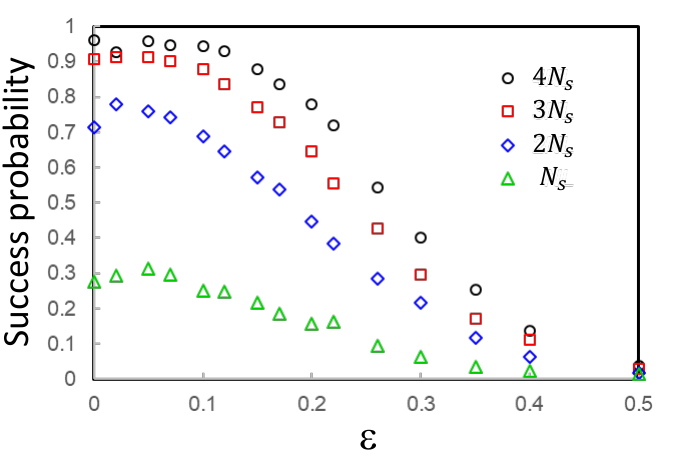}
\caption{The success probability for the MCMC-BF decoder as a function of the bit error rate $\varepsilon$ for the readouts of the MCMC sampler with four sample sizes. The sample size is expressed in a unit $N_s=\tbinom{K}{2}$, where $K=14$. 
\label{fig:12}
}
\end{figure}

We simulate noisy readout $\hat{\textbf{r}}'$ by multiplying the error matrix $\hat{\textbf{e}}$ generated by a pseudorandom number generator and the readout matrix $\hat{\textbf{r}}$, i.e.,  $\hat{\textbf{r}}'=\hat{\textbf{r}}\circ\hat{\textbf{e}}$ , where $\hat{\textbf{r}}$ has been obtained from the same spin-glass problem ($K_{14}$) as in the main text. Subsequently, we have applied the BF decoding to $\hat{\textbf{r}}'$, and investigated how the performance of the MCMC-BF hybrid decoder changes. To perform a statistical evaluation, we have run 1000 independent experiments. In Fig.\ref{fig:12}, we show the success probabilities for obtaining the correct state when incorporating IID errors into the readouts of the MCMC sampler with four sample sizes. In this figure, we have assumed that each spin flips with the bit error rate $\varepsilon$ $(0\leq\varepsilon\leq0.5)$, and that the Lagrange weights are the values at which the MCMC-BF hybrid decoder achieves maximum efficiency (corresponding to arrow \textit{B} in Fig.\ref{fig:9}). As expected, the probabilities reduce as $\varepsilon$ increases and reach almost zero at  $\varepsilon=0.5$. It has been confirmed that decoding succeeds even when $\varepsilon\neq0$, and, in particular, that the probability remains nearly unchanged in the region in which $\varepsilon\leq0.1$. This result suggests that the postreadout BF decoder can accommodate realistic noise in imperfect annealing devices.

\nocite{*}
\bibliography{APS}

\end{document}